\documentclass[aps,twocolumn,pra,notitlepage,]{revtex4-2}
\usepackage{amsmath,amssymb}
\usepackage{dsfont}
\usepackage{graphicx}
\usepackage{hyperref}
\usepackage{physics}
\usepackage{amsmath,amssymb}
\usepackage{dsfont}
\usepackage{graphicx}
\usepackage{hyperref}
\usepackage{physics}
\usepackage{mathtools}
\usepackage{cancel}
\usepackage{bbold}
\usepackage{bm}
\usepackage{color} 
\usepackage[dvipsnames]{xcolor}

\usepackage{soul}

\newcommand{\tun}{t_{c}}
\newcommand{\te}{\tau}  

\newcommand{\kb}{\mathbf k}

\begin{document}
\newcounter{theo}
\author{Jan A. Krzywda}\email{krzywda@ifpan.edu.pl}
\author{Łukasz Cywiński}
\affiliation{Institute of Physics, Polish Academy of Sciences, al.~Lotnik{\'o}w 32/46, PL 02-668 Warsaw, Poland}

\title{Interplay of charge noise and coupling to phonons in adiabatic electron transfer between quantum dots}

\begin{abstract}
Long-distance transfer of quantum information in  architectures based on quantum dot spin qubits will be necessary for their scalability. One way of achieving it is to simply move the electron between two quantum registers. Precise control over the  electron shuttling through a chain of tunnel-coupled quantum dots is possible when interdot energy detunings are changed adiabatically. Deterministic character of shuttling is however endangered by coupling of the transferred electron to thermal reservoirs: sources of fluctuations of electric fields, and lattice vibrations.
We theoretically analyse how the electron transfer between two  quantum dots is affected by  electron-phonon scattering, and interaction with sources of $1/f$ and Johnson charge noise in both detuning and tunnel coupling.
The electron-phonon scattering turns out to be irrelevant in Si quantum dots, 
while a
competition between the effects of charge noise and Landau-Zener effect leads to an existence of optimal detuning sweep rate, at which probability of leaving the electron behind is minimal. 
In GaAs quantum dots, on the other hand, coupling to phonons is strong enough to make the phonon-assisted processes of interdot transfer dominate over influence of charge noise. The probability of leaving the electron behind depends then monotonically on detuning sweep rate, and values much smaller than in silicon can be obtained for slow sweeps. 
However, after taking into account limitations on transfer time imposed by need for preservation of electron's spin coherence, minimal probabilities of leaving the electron behind in both GaAs- and Si-based double quantum dots turn out to be of the same order of magnitude. Bringing them down below $10^{-3}$ requires temperatures $\leq \! 100$ mK and tunnel couplings above $20$ $\mu$eV. 
\end{abstract}
\maketitle

\section{Introduction}
In quantum computing architectures based on voltage-controlled quantum dots (QDs), developed in GaAs/AlGaAs \cite{Hanson_RMP07,Volk_NPJQ19,Kuemmeth_roadmap_GaAs}, Si/SiGe \cite{Watson_Nature18,lawrieQuantumDotArrays2020a}, and silicon MOS \cite{Veldhorst_NC17,Huang_Nature19,Gonzalez_arXiv20,chanrionChargeDetectionArray2020} structures, scalability will be possible only if quantum information is transferred between few-qubit registers, separated by distances much larger than the typical QD size. This is caused by short-distance character of exchange interaction needed for two-qubit gates, and spatial extent of wiring needed for controlled application of voltages to the gates defining the dots, which together put limits on density of a qubit array \cite{Vandersypen_NPJQI17}. Coupling of electron spins to microwave photons is a possible mean of coherent coupling of spin qubits in GaAs \cite{scarlinoCoherentMicrowavephotonmediatedCoupling2019} and silicon \cite{Mi_Science17, Benito_PRB17,miCoherentSpinPhotonInterface2018,Samkharadze_Science18}. A conceptually simpler alternative, which has been recently pursued in experiments \cite{Hermelin_Nature11,McNeil_Nature11,Bertrand_NN16,takadaSounddrivenSingleelectronTransfer2019,mortemousqueEnhancedSpinCoherence2021,jadotDistantSpinEntanglement2020,Baart_NN16,Fujita_NPJQI17,Flentje_NC17,Mills_NC19,nakajimaCoherentTransferElectron2018c, vandiepenElectronCascadeDistant2021, yonedaCoherentSpinQubitc}, is to simply transfer an electron spin qubit over a large (at least a few micrometer) distance. 

We focus here on electron transfer along a chain of tunnel-coupled QDs \cite{Baart_NN16,Fujita_NPJQI17,Flentje_NC17,Mills_NC19,nakajimaCoherentTransferElectron2018c, vandiepenElectronCascadeDistant2021, yonedaCoherentSpinQubitc}. The shuttling is then caused by controlled tilting of energy levels of neighboring QDs that makes an electron move from one dot to the other. The basic step in such a process is single electron transfer between two tunnel-coupled QDs. In a simplified situation, in which we neglect spin and valley (in case of Si) degrees of freedom of the electron, the basic physics is captured by the Hamiltonian acting in a two-dimensional Hilbert space spanned by $\ket{L(R)}$ states, corresponding to electron localized in a local ground states of energy $E_{L(R)}$ in left (right) dot:
\begin{equation}
\label{eq:ham0}
\hat H = \frac{\epsilon }{2} \hat \sigma_z + \frac{\tun}{2}\hat \sigma_x \,\, , \
\end{equation}
where $\epsilon = E_L - E_R$ is the so-called interdot detuning of energy, $\tun$ is the tunnel coupling between the QDs, $\hat \sigma_z = \ketbra{L} - \ketbra{R}$, and $\hat \sigma_x = \ketbra{L}{R} + \ketbra{R}{L}$. For $\epsilon \! \ll \! -\tun$ the lowest energy state is localized in the $L$ dot, and this is the state that we take as an initial one in all the considerations below. 
For $\epsilon \! \gg \! \tun$ the lowest-energy state is localized in the $R$ dot, and one of course expects that for very slow change of $\epsilon$ from negative to positive values, the evolution will be adiabatic and the system will end up in this state. 
For a linear sweep, $\epsilon \! \propto \! v\te$, where $v$ is the rate of change of detuning, and constant $\tun$, we are dealing with classical Landau-Zener model \cite{Shevchenko_PR10}, for which the probability of having the electron in an exited state for $\epsilon(\te) \rightarrow \infty$ (i.e.~~leaving the electron behind in the L dot) is given by 
\begin{equation}
Q_{\text{LZ}} = \exp(-\pi \tun^2 /2 v) \,\, , \label{eq:QLZ}
\end{equation}
so that a near-perfect adiabatic transfer occurs when $\tun^2/v \! \gg \! 1$, i.e.~when the sweep rate is low. 

Changing the interdot detunings slowly is thus an obvious way to perform an on-demand deterministic transfer of an electron spin qubit. Of course, the total shuttling time should be much shorter than the spin coherence time of a moving electron, and according to Eq.~(\ref{eq:QLZ}) this requirement will put a lower bound on values of $\tun$ characterizing the chain of QDs. 
However, another issue needs to be addressed before we can claim to have a realistic estimate of sweep rate $v$ giving the smallest possible probability of error $Q$ in transfer between a pair of dots. Electrons are affected by charge noise unavoidable in semiconductor nanostructures, and coupled to lattice vibrations. As we show in this paper, interactions with sources of electric field noise and phonons in realistic Si- and GaAs-based structures are dominating the physics of charge transfer in a wide range of sweep rates, with nonadiabatic effects described by Landau-Zener theory being relevant only for very fast sweeps. 

Since our focus here is on open system character of an electron tunneling between two quantum dots, we use the above-described simplest possible two-level model of the closed system. Taking into account the spin degree of freedom and spin-orbit cupling that affects its dynamics during the electron motion in GaAs \cite{Li_PRA17} (and to a smaller extent in silicon \cite{Li_PRA17,Ginzel_PRB20}), and then a valley degree of freedom in Si \cite{Friesen_PRB07,Culcer_PRB10,Zwanenburg_RMP13}, leads to 4- or 8-level models with multiple anticrossings of states
\cite{Li_PRA17,Zhao_SR18,Zhao2018b, Cota_JPCM18, Shevchenko_PRB18, Krzywda_PRB20,Ginzel_PRB20, Buonascorsi_PRB20, mallaNonadiabaticTransitionsLandauZener2021}. 
The two-level model used here exhibits a simpler behavior in closed system case, and using it will typically lead to an underestimation of unwanted effects due to not-slow-enough sweeps (for a closed system), and coupling to environment (for an open system). The results given in this paper consequently correspond to the best-case scenario for given $\tun$ and assumed magnitudes of charge noise and temperature.

The physics of Landau-Zener effect in presence of coupling to environment has obviously been a subject of multiple works. Dissipative adiabatic evolution affected by coupling to bosonic baths having Ohmic spectrum was most often considered \cite{yamaguchiMarkovianQuantumMaster2017, arceciDissipativeLandauZenerProblem2017, leggettDynamicsDissipativeTwostate1987,  javanbakhtDissipativeLandauZenerQuantum2015, huangDynamicsDissipativeLandauZener2018, zuecoLandauZenerTunnelling2008}. It is known \cite{saitoDissipativeLandauZenerTransitions2007, wubsGaugingQuantumHeat2006} that coupling to zero-temperature bath suppresses the final occupation of the higher-energy state (``the electron being left behind in the initial dot'' in the physical scenario of interest here), while at finite temperature this occupation can be enhanced \cite{Ao_PRB91,pokrovskyFastQuantumNoise2007,Kayanuma_PRB98,arceciDissipativeLandauZenerProblem2017,nalbachAdiabaticMarkovianBathDynamics2014}. 
Such effects of coupling to low-temperature reservoirs were discussed in many physical contexts \cite{bensenyAllWellThat2020, yamaguchiMarkovianQuantumMaster2017,chenWhyWhenPausing2020}.
Stochastic modifications of LZ parameters were also considered \cite{Kayanuma_JPSJ84,Kayanuma_JPSJ85, dodinLandauZenerTransitionsMediated2014}, including fast classical fluctuations \cite{Pokrovsky_PRB03} and noise characterized by non-trivial spectral density \cite{Luo_PRB17,Malla_PRB17,Sinitsyn_PRB03,vestgardenNonlinearlyDrivenLandauZener2008,Krzywda_PRB20}, including 1/f type noise, the tail of which also resulted in incoherent transitions between the states \cite{aminMacroscopicResonantTunneling2008,Krzywda_PRB20,youPositiveNegativefrequencyNoise2021}. 
In this paper we focus on quantum dots based on silicon and GaAs, and employ realistic models of charge noise (having both Johnson/Ohmic and $1/f$ type spectra, and coupling to both $\epsilon$ and $\tun$), and phonon interaction with an electron confined in a double quantum dot. We use the Adiabatic Master Equation \cite{albashQuantumAdiabaticMarkovian2012, nalbachAdiabaticMarkovianBathDynamics2014, yamaguchiMarkovianQuantumMaster2017}, in which the influence of the environment (actually a few distinct reservoirs in the case discussed here) is modeled with energy-dependent rates of transitions between instantaneous eigenstates of the slowly changing Hamiltonian of the system. For negligible probability of coherent Landau-Zener excitation, this approached reduces to a simple differential rate equation \cite{vogelsbergerButterflyHysteresisCurve2005,haikkaDissipativeLandauZenerLevel2014}), which we solve in a way analogous to the one described in \cite{nalbachAdiabaticMarkovianBathDynamics2014}. 

During the detuning sweep, the energy gap between eigenstates of instantaneous Hamiltonian varies between $\tun \sim$ $10$ $\mu$eV and largest value of $\epsilon \! \sim \! 1$ meV. With temperatures in experiments typically around $100$ mK, corresponding to thermal energy of $\approx \! 10$ $\mu$eV, we should expect a nontrivial role of temperature dependence of rates of energy absorption and emission by the reservoirs. Note that in our previous work \cite{Krzywda_PRB20} we have focused on influence of {\it classical} (i.e.~high-temperature) $1/f$ charge noise on electron transfer. Here we address the situation of lower temperatures/larger tunnel couplings, taking into account the quantum limit \cite{Clerk_RMP10} of both $1/f$ noise from two-level fluctuators present in the nanostructure, and Johnson noise from reservoirs of free electrons, while furthermore considering the coupling of the moving electron to phonons. Coupling to all these thermal reservoirs gives transition rates, $\Gamma_{+/-}(\Omega)$ for transfer of energy $\Omega$ from/to the environment, that nontrivially depend on $\Omega$. The detailed balance between them, which reads $\Gamma_{+}/\Gamma_{-} \! =\! e^{-\beta \Omega}$, has the following general consequence for the dynamics of the system. With the system initially in ground state, transitions into an excited are exponentially suppressed for large negative detunings, and they start to become increasingly efficient as we approach the anticrossing of levels, at which the gap is minimal and equal to $\tun$. This effect of enhancement of excitation rate at the anticrossing is additionally strengthened in the considered system by the fact that an electron delocalized between the two dots is more susceptible to both charge noise and interaction with phonons (as the transitions between states localized in each dot that govern the dynamics in far-detuned regimes are suppressed by small overlap of wavefunctions). The finite occupation of the ``wrong'' dot generated during passing through $|\epsilon| \! \lesssim \! \tun$ region can then be diminished (``healed'' in the terminology used below) by processes of energy emission into the reservoirs that dominate over processes of energy absorption by them when $\epsilon \! \gg \! k_{B}T$. Arriving at  the final result of interplay between environment-induced excitation near the anticrossing, and the subsequent energy relaxation (the environment-assisted dissipative tunneling into the ``correct'' final state), requires consideration of realistic coupling to all the reservoirs at temperatures and sweep rates relevant for experiments in quantum dots. Such a careful consideration is the goal of this paper.  

Our key qualitative result concerning application to realistic quantum dots, is that in Si-based structures (both Si/SiGe and SiMOS) the dominant process disturbing the adiabatic evolution close to anticrossing of levels is due to charge noise (with coupling to phonons giving transition rates $3$ order of magnitude smaller than those estimated for charge noise), and the finite probability of leaving the electron behind is subsequently diminished by relaxation processes due to charge noise and phonons that occur at large detunings only when the transfer is very slow. 
On the other hand, in GaAs/AlGaAs structures the piezoelectric coupling to phonons dominates over coupling to charge noise over a wide range of detunings, and consequently the processes involving energy exchange between the transferred electron and lattice vibrations dominate the physics of the problem. The longer the charge transfer takes, the more time the system spends in far-detuned regime in which the energy gap exceeds thermal energy, and the closer it gets to a thermalized state characterized by small occupation of higher-energy level, i.e.~of the electron being in the wrong dot. Phonons thus help in maintaining a deterministic character of the charge transfer. These conclusions are quite robust against modifications of parameters of high-frequency properties of Johnson and $1/f$ type charge noises considered here.  

The article is organized in the following way, in Sec.~\ref{sec:model} we set up the problem for the closed system and discuss the adiabatic condition for its dynamics, introduce the Adiabatic Master Equation as an approach to open system dynamics, and discuss a few physically transparent (and, as we show later, relevant for the case of electron transfer in silicon- and GaAs-based quantum dots) approximate solutions of this equation. In Section \ref{sec:rates} we calculate the detuning-dependent transition rates between instantaneous eigenstates of the two-level Hamiltonian.
We perform calculations for coupling  to phonons, and finite-temperature environments that cause charge noise of both Johnson and $1/f$ type in detuning and tunnel coupling. We give there a discussion of expected amplitude of $1/f$ noise at GHz frequencies relevant for transitions during electron transfer in realistic GaAs- and silicon-based quantum dots. Finally, in Section \ref{sec:results} we use these rates to calculate the dynamics of the electron driven adiabatically through an anticrossing of levels associated with the two dots, and show a qualitative difference between resulting probability of ``leaving the electron behind'' between GaAs- and silicon-based quantum dots. In the last Section we discuss some of the implications of these results for experimental efforts aimed at using chains of quantum dots for coherent shuttling of electron spin qubits.

\section{Model of System's dynamics}  \label{sec:model}

\subsection{Adiabatic condition for closed system}
We consider two energy levels that in the double quantum dot case correspond to the lowest-energy orbital states localized in each of the two dots, $\ket{L}$ and $\ket{R}$. In case of silicon QDs we assume that the valley splitting is large enough for us to consider a single anticrossing of two lowest-energy valley-orbital levels. We also neglect the spin degree of freedom - interplay between the nonadiabatic effects in charge transfer and dynamics of the spin of the transferred electron will be discussed elsewhere \cite{SiQuBus}. We therefore work with the model defined by Hamiltonian from Eq.~\eqref{eq:ham0}, in which we now assume that $\epsilon$ and $\tun$ depend on time $\tau$.

For any value of $\epsilon(\tau)$ and tunnel coupling $\tun(\tau)$, the Hamiltonian $\hat{H}(\tau)$ has eigenstates  
\begin{align}
\label{eq:states}
\ket{+,\theta(\tau)} & = \cos[\theta(\tau)/2]\ket{R} + \sin[\theta(\tau)/2]\ket{L} \nonumber \\
\ket{-,\theta(\tau)} & = \cos[\theta(\tau)/2]\ket{L} - \sin[\theta(\tau)/2]\ket{R},
\end{align}
where $\theta(\tau) = \arccot(-\epsilon(\tau)/|\tun(\tau)|)$. The discussion of nonadiabatic effects due to time-dependence of $\epsilon$ and $\tun$, or effects of interaction with the environment, is most transparent if we transform the state of the system into an ``adiabatic frame'' \cite{demirplakAdiabaticPopulationTransfer2003}: instead of working with $\ket{\psi(\tau)}$ which fulfills $i\partial_{\tau} \ket{\psi(\tau)} = \hat{H}(\tau)\ket{\psi(\tau)}$ we work with $|\tilde{\psi}(\tau)\rangle \! \equiv \! \hat U[\theta(\tau)]  \ket{\psi(\tau)}$, where a time-dependent  unitary operator 
\begin{equation}
\hat U[\theta(\tau)] = \exp[i\hat \sigma_y \theta(\tau)/2] \,\, .
\end{equation}
transforms the $R/L$ states into the instantaneous eigenstates of $\hat{H}(\tau)$: $\ket{+(-),\theta(\tau)} = \hat U[\theta(\tau)] \ket{R(L)}$. One can see that for a perfectly adiabatic evolution of the system, for which an initial superposition of eigenstates of $\hat{H}(\tau_i)$ at given $\tau_i$ evolves into {\it the same} superposition of eigenstates of $\hat{H}(\tau_f)$ at the final time $\tau_f$, the transformed
$|\tilde{\psi}(\tau)\rangle$ state  is {\it time-independent}. Indeed, the evolution in the adiabatic frame is controlled by 
\begin{equation}
\mathcal{H}(\tau)  =  \hat{U}[\theta(\tau)]\hat{H}(\tau)\hat{U}^{\dagger}[\theta(\tau)] - i  \hat{U}[\theta(\tau)]\left (\frac{\partial \hat{U}^\dagger[\theta(\tau)]}{\partial \tau} \right) \,\, ,
\end{equation}
which for the system discussed here reads
\begin{equation}
\label{eq:ham_ad}
\hat {\mathcal{H}}(\tau) =\frac{\Omega(\tau)}{2}\hat \varsigma_z - \frac{\dot \theta(\tau)}{2}\hat \varsigma_y,
\end{equation}
where $\hat \varsigma_z$, $\hat \varsigma_y$ are Pauli operators in $\ket{+,\theta(\tau)},\ket{-,\theta(\tau)}$ basis of instantaneous eigenstates of the time-dependent Hamiltonian $\hat{H}(\tau)$, $\dot \theta \! =\! \mathrm{d}\theta(\tau)/\mathrm{d}\tau$, and the instantaneous energy splitting is
\begin{equation}
    \Omega(\tau) = \sqrt{\epsilon^2(\tau) + \tun^2(\tau)} \,\,. \label{eq:Omega}
\end{equation}

We assume the electron is initialized in the ground state at large negative detuning $\epsilon(-\tau_\infty)\ll t_c$, such that the initial state $\ket{\psi(-\tau_\infty)}=\ket{-,\theta(-\tau_\infty)}\! \approx \!\ket{L}$. 
Due to non-negligible coupling between the adiabatic states during the system's evolution (i.e.~a nonzero $\dot\theta(\tau)$ term in Eq.~\eqref{eq:ham_ad}), a non-zero occupation of excited state $\ket{+,\theta(\tau)}$ can be generated. When the detuning sweep terminates at large $\epsilon(\tau_\infty)\gg t_c$, the occupation of excited state defines the transfer error, i.e.~the probability of the electron being left behind in the $L$ dot:
\begin{equation}
Q = |\bra{\psi(\tau_\infty)}\ket{+,\theta(\tau_\infty)}|^2 \approx |\braket{\psi(\tau_\infty)}{L}|^2.
\end{equation}
The calculation of $Q$ for an electron coupled to environments relevant for semiconductor-based gated quantum dots is the main goal of this paper.

For constant tunnel coupling $\tun$, and for $\epsilon(\tau) = v\tau$ we are dealing with the well-known Landau-Zener model \cite{shevchenkoLandauZenerStuckelberg2010}, in which $Q$ is given by $Q_{LZ}$ from Eq.~(\ref{eq:QLZ}). We concentrate here on the adiabatic regime, defined by $Q_{\text{LZ}}\! \ll \! 1$, which implies $\tun^2/ v \! \gg \! 1$, and means that the ratio of ``transverse'' and ``longitudinal'' terms in the effective Hamiltonian, Eq.~(\ref{eq:ham_ad}), fulfills
\begin{equation}
\label{eq:adiab_cond}
\frac{\dot\theta}{\Omega} = \frac{v \tun}{\Omega^3} < \frac{v}{\tun^2}  \ll 1 \,\, .
\end{equation} 
This is the adiabatic condition for the dynamics of closed and noise-free system. When it is fulfilled during detuning sweep, the electron remains at all times in the ground state $\ket{-,\theta(\tau)}$, which means it physically moves from the state initially localized in the left dot $\ket{-,\theta(-\tau_\infty)} \! =\!  \ket{L}$, to a final state $\ket{-,\theta(\tau_\infty)) } \! = \! \ket{R}$, located in the right dot.

\subsection{Dynamics of an open system}
We use here Adiabatic Master Equation (AME) approach \cite{pokrovskyFastQuantumNoise2007,albashQuantumAdiabaticMarkovian2012,nalbachAdiabaticMarkovianBathDynamics2014,chenWhyWhenPausing2020}, in which transitions caused by the environment occur between the instantaneous eigenstates of $\hat{H}(\tau)$, which are given by Eqs.~\eqref{eq:states}.  
Our focus on the adiabatic regime ($t_c^2\gg v$), combined with relatively weak coupling to charge noise (with noise RMS $\sigma \ll t_c$) and short intrinsic correlation time of phonon bath allow us to use here a Lindbladian form of AME  \cite{arceciDissipativeLandauZenerProblem2017,yamaguchiMarkovianQuantumMaster2017}, which reads:  
\begin{equation}
\label{eq:rate_H0}
\partial_\tau {\hat\varrho}_A = i [\hat{\mathcal{H}}(\tau),\hat \varrho_A] +  \Gamma_+(\tau) \mathcal D[\hat \varsigma_+]\hat \varrho_A + \Gamma_-(\tau) \mathcal D[\hat \varsigma_-]\hat \varrho_A,
\end{equation}
where $\hat \varrho_A = \hat U_y(\theta) \hat \varrho_{\text{LR}}\hat U_y^\dagger(\theta)$ and $\varrho_{\text{LR}}$ is the density matrix of the system before switching the description to the ``adiabatic'' frame, $\hat \varsigma_+ =\ketbra{+,\theta}{-,\theta}$, $\hat \varsigma_- =\ketbra{-,\theta}{+,\theta}$, and
\begin{equation}
\mathcal{D}[\hat o]\hat \varrho = \hat o \hat\varrho o^\dagger - \frac{1}{2}\acomm{\hat o^\dagger \hat o}{ \hat \varrho}
\end{equation}
is the Linbladian associated with operator $\hat o$ and time-dependent relaxation/excitation rate $\Gamma_{\pm}(\tau)$.  In this approach these rates depend on time though their dependence on the value of instantaneous energy splitting $\Omega(\tau)$ from Eq.~(\ref{eq:Omega}), 
i.e.~$\Gamma_{\pm}(\tau) \! =\! \Gamma_{\pm}[\Omega(\tau)]$. Below we will use both notations, $\Gamma_{\pm}(\tau)$ and $\Gamma_{\pm}[\Omega(\tau)]$, depending on context.
In particular, if noise-induced excitations dominate over the Landau-Zener effect due to deterministic time-dependence of $\hat{H}(\tau)$, i.e.~$Q_\text{noise} \gg Q_{\text{LZ}}$, the unitary evolution can be safely neglected and Eq.~\eqref{eq:rate_H0} reduces to a simple rate equation
\begin{equation}
\label{eq:rate}
\dot Q(\tau) \approx \Gamma_+(\tau) - Q(\tau)\big(\Gamma_-(\tau)+\Gamma_+(\tau)\big),
\end{equation}
where $Q(\tau) = \bra{+,\theta}\hat \varrho_A(\tau) \ket{+,\theta}$ denotes occupation of the higher energy state $\ket{+,\theta}$ at time $\tau$.

Given the initial condition $Q(-\tau_\infty) = 0$, the solution to Eq.~\eqref{eq:rate} reads
\begin{align}
\label{eq:Q}
Q &= \int_{-\tau_\infty}^{\tau_\infty} \text{d}\tau \,\Gamma_+(\tau) e^{-\int_{\tau}^{\tau_\infty} \Gamma_+(\tau') + \Gamma_-(\tau')\text{d}\tau'}=\nonumber \\
 &= \int_{-\tau_\infty}^{\tau_\infty}\text{d}\tau \Gamma_+(\tau) e^{-\chi(\tau,\tau_\infty)}.
\end{align}
where
\begin{equation}
\label{eq:hf}
\chi(\tau, \tau_\infty) = \int_{\tau}^{\tau_\infty} \left( \Gamma_+ + \Gamma_-\right) \text{d} \tau' \,\, .
\end{equation}

\subsection{Approximate solutions} \label{sec:approximate}
Let us now discuss a few physically motivated approximate solutions for the probability of ending up in the excited state at the end of the sweep $Q$, i.e. the probability that the electron remains in the initial dot. We start with a simplest perturbative approach to rate equation \eqref{eq:rate}, assuming $\Gamma_\pm \tau_{\infty} \! \ll \! 1$. In the lowest order one can write:
\begin{equation}
\label{eq:q1}
    Q^{(1)} = \int_{-\tau_\infty}^{\tau_\infty}\Gamma_+(\tau) \text{d}\tau,
\end{equation}
As the energy needed for transition from ground to excited state comes from thermal fluctuations of environment, the excitation rate $\Gamma_+$ is strongly suppressed at low temperatures, when $k_{B}T \! \ll \! \tun \! \leq \Omega(\tau)$. At these temperatures the rate of energy relaxation into the environment, $\Gamma_{-}(\Omega)$, is temperature-independent, as the thermal occupation factor for environmental states of energy $\Omega \! \gg \! k_B T$ is zero, and $\Gamma_{-}(\Omega)$ depends then only on density of environmental states and coupling matrix elements. For all the environments considered in this paper, these dependencies lead to a power-law behavior of the rates, $\Gamma_{-}(\Omega) \! \propto \! \Omega^{a}$ with $a\! \in \! [-3,3]$ depending on  the transition mechanism and range of $\Omega$, see derivations in the next Section. 
As we assume the environment to be in thermal equlibrium, the detailed balance condition, which reads $\Gamma_+(\Omega) = \Gamma_-(\Omega) e^{-\beta \Omega}$, leads to $\Gamma_+(\Omega) \propto \Omega^a e^{-\beta \Omega}$ with $e^{-\beta \Omega} \! \ll \! 1$ at low temperatures.

The excitation process takes then place in a narrow range of detunings around the avoided crossing, as $\Gamma_{+}(\Omega)$ very quickly decreases when $|\epsilon|$ increases. As $\Omega(\tau) \approx \beta t_c + \beta v^2 \tau^2/2t_c$ for $\epsilon \! \ll \! \tun$, we neglect in this regime the $\epsilon$ dependence of $\Gamma_{+}(\Omega)$ and replace it with value for $\Omega \! =\! \tun$  (equivalently: for $\tau\! =\! 0$), while we keep it in the thermal factor. The integrand in \eqref{eq:q1} can then be approximated as $\Gamma_+(\tau) \approx \Gamma_-(0) e^{-\beta \Omega(\tau)}$, and the integration can be done over a range of $|\epsilon| \! \ll \! \tun$. In this way we obtain the Single Excitation Approximation Limit (SEAL):
\begin{align}
     Q_{\text{SEAL}} &  =  \Gamma_-(0) e^{-\beta t_c} \int_{-\infty}^{\infty}  e^{-\frac{\beta v^2 t^2}{2t_c}}\text{d}\tau \,\, ,\nonumber\\
     & = \frac{\sqrt{2\pi k_{\text{B}}T t_c}}{v}  \Gamma_-(0) e^{-\beta t_c}\,\, ,  \label{eq:seal}
 \end{align}
 which assumes that at most a single quantum jump from ground to excited state takes place in the avoided crossing region.

The SEAL approximation does not take into account possibility of electron transition in the opposite direction, i.e. from excited to ground state, which would lead to partial recovery of ground state occupation - an effect that we will refer to  as  a ``healing'' of excitation that occurred close to the anticrossing. 
This effect is captured by the $\exp[-\chi(\tau)]$ factor in Eq.~\eqref{eq:Q} with  $\chi(\tau)$, given in Eq.~\eqref{eq:hf}),  evaluated in the low-temperature limit of $\Gamma_{-} \! \gg \! \Gamma_+$. 
The effect of transitions occuring during the part of the sweep when $\epsilon(\tau) \! > \! \tun$ is captured by a \textit{Healed Excitation Approximation Limit} (HEAL):
\begin{equation}
\label{eq:q1chi}
Q_{\text{HEAL}} \approx Q_{\text{SEAL}} \exp{-\int_{0}^{\tau_\infty} \Gamma_-(\tau) \,\text{d}\tau}\, .
\end{equation}  

\begin{figure}[htb!]
\includegraphics[width=\columnwidth]{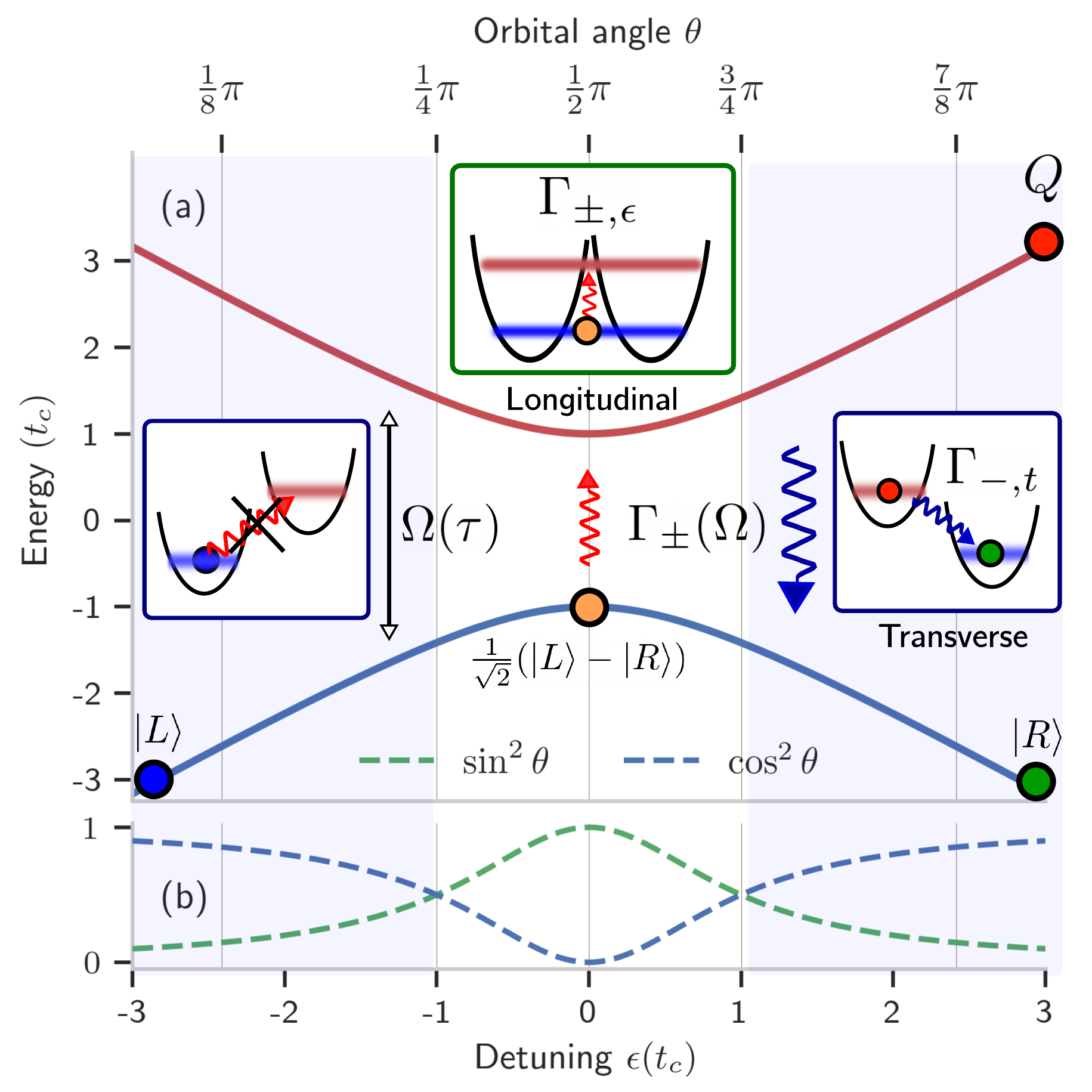}
\caption{A schematic picture of adiabatic transition between two quantum dots in presence of charge noise and phonon bath. In panel (a) we show energy of instantaneous states $\ket{\pm,\theta}$(red/blue lines) as a function of detuning $\epsilon$ (lower axis) and orbital angle $\theta = \acot(-\epsilon/t_c)$ (upper axis). Despite detuning sweep being adiabatic ($t_c^2\gg v$), the electron initialized in the left dot (blue circle) can still end up with non-zero occupation of excited state localized in right dot $Q$ (red circle), as a result of coupling to environment, which leads to incoherent transitions between eigenstates of the instantaneous Hamiltonian characterized by the rates $\Gamma_{\pm}(\Omega)$. At low temperatures, the excitation rate is non-negligible only in vicinity of avoided crossing, where the gap is smallest, $\Omega(0) = t_c$, while relaxation accounts for recovery of ground state occupation (the ``healing'' of the excitation at larger detuning. As the detuning is changed, the character of eigenstates of the instantaneous Hamiltonia, $\ket{\pm,\theta}$, is modified from dot-like character at $|\epsilon| \gg t_c$ to orbital-molecular-like at $|\epsilon|\ll t_c$, which is accompanied by dominant role of transverse (interdot) and longitudinal transitions, respectively, see Sec.~\ref{sec:rates_general}. To illustrate the difference between them, in panel (b) we plot $\cos^2\theta = |\bra{+,\theta}\sigma_z\ket{+,\theta}|^2$ (dashed blue) and $\sin^2\theta = |\bra{+,\theta}\sigma_x\ket{+,\theta}|^2$ (dashed green) factor that determine the relative importance of $\Gamma_t$ and longitudinal $\Gamma_\epsilon$ relaxation channels, respectively. Insets with green and blue frames schematically representing transition mechanisms dominant in regimes of $\epsilon \! \ll \! -\tun$, $|\epsilon| \! \ll \! \tun $, and $\epsilon \! \gg \! \tun$. 
}
\label{fig:schematic}
\end{figure}

The physical picture expected to hold at low $T$ is thus the following. A finite $Q$ is generated due to coupling to a thermal reservoir near the anticrossing, and then processes of emission of energy into this reservoir lead to a diminishing of its final value at the end of the sweep,  making the final state of the system closer to the one following from an ideal adiabtic evolution. Such a healing process results in environment-assisted inelastic tunneling into ground state at the end of the driving, see Fig.~\ref{fig:schematic}.
In Sec.~\ref{sec:results} we will demonstrate in which regimes of parameters the SEAL/HEAL solutions are applicable for realistic DQD devices.

Note that up to this moment we have not specified any particular form of relaxation/excitation rates $\Gamma_{\pm}(\tau)$, which makes above approximations suitable also for other systems described in terms of the L-Z Hamiltonian \eqref{eq:ham0}, in the adiabatic limit ($t_c^2\gg v$) and coupled to environment at relatively low-temperature ($t_c \gtrsim k_{\text{B}}T$). 

\section{Transition rates for an adiabatically transferred electron}
\label{sec:rates}
\subsection{General properties} \label{sec:rates_general}
We consider now a transfer of an electron between two quantum dots that is driven by a detuning sweep slow enough to be adiabatic in the closed system limit. After turning on a weak coupling to an environment, the transition rates $\Gamma_{\pm}(\tau)$ in the Adiabatic Master Equation (AME) from Eq.~\eqref{eq:rate_H0} are evaluated at given $\tau$ as if the system described by the instantaneous Hamiltonian $\hat{H}(\tau)$ from Eq.~\eqref{eq:ham0}, was subjected to an off-diagonal coupling with an environment for a long enough time for Fermi Golden Rule (FGR) calculation to be applicable. Thus the general form of electron-environment coupling $\hat V = \tfrac{1}{2}(\hat V_t \hat \sigma_x + \hat V_\epsilon \hat \sigma_z)$ in the $\ket{L/R}$ basis, at given $\theta(\tau)$ should be expressed in the basis of eigenstates of instantaneous Hamiltonian, $\ket{\pm,\theta}$, using $\hat{\mathcal V}(\theta) = \hat U_y^\dagger(\theta) \hat V \hat U_y(\theta)$, which leads to
\begin{equation}
\label{eq:adiab_v}
    \hat{\mathcal{V}}(\theta) = \tfrac{1}{2}(\hat{\mathcal{V}}_x \,\hat{\varsigma}_x + \hat{\mathcal{V}}_z \,\hat \varsigma_z),
\end{equation}
where $\hat {\mathcal V}_x = (\hat{V}_t \cos\theta + \hat{V}_\epsilon \sin\theta)$ and $\hat {\mathcal V}_z = (\hat{V}_\epsilon \cos\theta - \hat{V}_t \sin\theta)$.
This means that at every $\tau$ we do the FGR calculation for $\hat{\mathcal V}_x \hat{\varsigma}_x/2$ coupling, where $\hat{\varsigma}_x$ acts in basis of eigenstates of the instantaneous $\hat{H}(\tau)$. With the environmental Hamiltonian given by $\hat{H}_E$, we calculate then the quantum spectral density for the operator $\hat{\mathcal V}_x(t) = e^{i\hat{H}_{E}t}\hat{\mathcal V}_xe^{-i\hat{H}_{E}t}$, given by \cite{Schoelkopf_spectrometer,Clerk_RMP10}
\begin{equation}
\label{eq:qspectr}
    S^{Q}_{\mathcal{V}}(\omega) = \int_{-\infty}^{\infty} \langle \hat{\mathcal V}_x(t)\hat{\mathcal V}_x(0)\rangle e^{i\omega t} \mathrm{d}t
\end{equation}
where $\langle \ldots \rangle \! =\! \mathrm{Tr}_{E}(\hat{\rho}_{E}\ldots )$ is the averaging over the environmental density matrix $\hat{\rho}_{E}$. The rate of excitation of the system, i.e.~a transition that involves taking energy $\Omega(\tau)$ from the environment, is then given by \cite{Schoelkopf_spectrometer,Clerk_RMP10}
\begin{equation}
\Gamma_{+}(\tau) = \frac{1}{4}S^{Q}_{\mathcal{V}}[-\Omega(\tau)] \,\, , \label{eq:Gamma_minus}
\end{equation}
while the relaxation rate is 
\begin{equation}
\Gamma_{-}(\tau) = \frac{1}{4}S^{Q}_{\mathcal{V}}[\Omega(\tau)] \,\, . \label{eq:Gamma_plus}
\end{equation}
For an environment in thermal equilibrium which we consider here, we have $\hat{\rho}_{E} \! \propto \! e^{-\beta\hat{H}_{E}}$ and the detailed balance condition, $S^{Q}_{V}(\Omega) \! =\! S^{Q}_{V}(-\Omega)e^{\beta\Omega}$, and thus $\Gamma_{-}[\Omega(\tau)]\! = \! \Gamma_{+}[\Omega(\tau)]e^{\beta\Omega(\tau)}$, is fulfilled. 

As the longitudinal $\hat V_\epsilon$ and transverse $\hat V_t$ couplings in dots basis are often of different physical origin, we assume $\langle \hat V_\epsilon \hat V_t \rangle = 0$, so that the transition rate can be written as $\Gamma_\pm(\tau) = \Gamma_{\epsilon,\pm}(\tau) + \Gamma_{t,\pm}(\tau)$, where we introduced 
\begin{align}
\label{eq:long_tans}
    \text{(Longitudinal) :} \quad \Gamma_{\epsilon,\pm}(\tau) &= \frac{1}{4} \sin^2\theta\,\, S^Q_\epsilon[\mp \Omega(\tau)],\, \nonumber \\
    \text{(Transverse) :} \quad \Gamma_{t,\pm}(\tau) &= \frac{1}{4} \cos^2\theta\,\, S^Q_t[\mp \Omega(\tau)]
\end{align}
contributions, defined using spectral densities of $\hat{V}_{\epsilon}$ and $\hat{V}_t$ operators, $S_\epsilon^Q(\omega) = \int \langle \hat V_\epsilon(t)\hat V_\epsilon(0) \rangle e^{i\omega t}\text{d}t$ and $S_t^Q(\omega) = \int \langle V_t(t) \hat V_t(0) \rangle e^{i\omega t}\text{d}t$. The $\hat{V}_\epsilon \hat{\sigma}_z$ coupling that is longitudinal in the $\ket{L/R}$ basis (the dot basis) appears due to fluctuations of detuning or phonons coupling to the operator $\hat \sigma_z$. It is most efficient at causing transitions between $\ket{\pm,\theta}$ states when the latter have a molecular-orbital character, i.e.~$\theta \! \approx \! \pi/2$, $\sin^2\theta \! \approx \! 1$, $\ket{\pm,\pi/2} =\tfrac{1}{2}( \ket{L} \pm \ket{R})$, and $|\epsilon| \! \ll \! \tun$.  On the other hand, the  transverse coupling $\hat{V}_t \hat{\sigma}_x $ is due to fluctuations of tunnel coupling or phonons coupling to the operator $\hat \sigma_x$. It leads to transitions of interdot character between the states $\ket{L} \leftrightarrow \ket{R}$ that correspond to $\ket{\pm,\theta}$ states at $\theta \! \ll \! 1$ and $\theta \! \approx \! \pi$ (i.e.~$\epsilon \! \ll \! -\tun$ and $\epsilon \! \gg \! \tun$, respectively), see Fig.~\ref{fig:SiGe_GaAs_tc}b. 
 Below we will see that for all the considered mechanisms, the transverse processes are weaker than the longitudinal ones, i.e. $S_t^Q(\Omega) \ll S_\epsilon^Q(\Omega)$, so the latter could become relevant only in a very far-detuned regime. 

For an electron in a double quantum dot, the relevant mechanisms of transitions between the eigenstates are due to coupling of electron charge to two reservoirs: lattice vibrations (phonons) and sources of fluctuations of electric fields -- free electrons in metallic electrodes and ungated regions of semiconductor quantum well being the sources of Johnson noise, and bound charges switching between a discrete number of states being the sources of $1/f$ type noise \cite{Paladino_RMP14}.
Due to their distinct physical origin, we neglect correlations between different transition mechanisms and write the relaxation rate as:
\begin{equation}
\label{eq:gamm}
\Gamma_- = \Gamma_-^{(\text{ph})}+ \Gamma_-^{(\text{1/f})} + \Gamma_-^{(\text{Joh})}.
\end{equation}
In the above, we separated charge noise contribution into $\Gamma^{(\text{1/f})}$ due to tail of 1/f-like noise from two-level fluctuators \cite{youPositiveNegativefrequencyNoise2021} in the quantum well interface and $\Gamma^{(\text{Joh})}$ due to Johnson's noise caused by wiring in vicinity of quantum dots \cite{marquardtSpinRelaxationQuantum2005}. As a direct consequence of Eq.~\eqref{eq:gamm} the exact formula for leaving electron in the initial dot reads:
\begin{align}
Q = \int_{-\tau_\infty}^{\tau_\infty}\text{d}\tau \sum_m \Gamma_+^{(m)}(\tau) \exp(-\sum_{m'}\chi^{(m')}(\tau,\tau_\infty)),
\end{align}
where indices $m,m'$ stands for phonon, 1/f or Johnson's mechanisms, while $\chi^{(m)}(a,b) = \int_a^b \Gamma_+^{(m)}(\tau')+\Gamma_-^{(m)}(\tau') \text{d}\tau'$.

Let us discuss now the quantum noise spectra relevant for the two types of reservoirs being the sources of charge noise, and the lattice vibrations. 

\subsection{Charge noise}
\label{sec:charge_noise}
The way in which sources of charge noise couple to the electron in a DQD is most easily visible if we consider the high-temperature (or low energy transfer) limit of $\beta \Omega \! \ll \! 1$. The quantum spectral density becomes then symmetric in frequency, $S^{Q}_{V}(\Omega) \! =\!S^{Q}_{V}(-\Omega)$ (so that $\Gamma_{+} \! =\! \Gamma_{-}$), and it can be identified with a classical power spectral density of a classical stochastic process describing the fluctuations of the electric fields caused by the dynamics of the reservoir. These processes manifest themselves as time-dependent corrections to parameters of $\hat{\mathcal{H}}(\tau)$:  $\delta \epsilon(\tau)$ and $\delta t(\tau)$ for detuning and tunnel coupling noise, respectively. As long as the amplitude of the noise is small ($\delta t, \delta \epsilon \ll t_c$), the modification of the instantaneous splitting $\Omega(\tau) = \sqrt{(v\tau +\delta \epsilon)^2 + (t_c+ \delta t)^2}$ is negligible. However, time variation of $\delta \epsilon$ and $\delta t$ activates coupling between the eigenstates of the instantaneous Hamiltonian from Eq.~\eqref{eq:ham_ad}), as in the lowest order in $\delta \epsilon$ and $\delta t$ we have
\begin{align}
\label{eq:dth_der}
\dot \theta&= \frac{\partial}{\partial \tau} \acot(-\frac{v\tau+\delta \epsilon}{t_c + \delta t}) \approx  \frac{\sin\theta{\delta \dot \epsilon}  + \cos\theta {\delta \dot t}}{\Omega_0} \,\, ,
\end{align} 
where $\Omega_0 \! =\! \sqrt{v^2\tau^2 +\tun^2}$, and the last approximation relies on $t_c^2\gg v$ assumption to neglect contributions not larger than the noiseless coupling $\dot \theta_0 = v t_c/\Omega_0^2 \ll \Omega$, see Eq.~\eqref{eq:adiab_cond}. As we neglect correlations between $\delta \epsilon $ and $\delta t$, we treat the transitions induced by these two noises independently. Taking then into account that the classical spectrum $S^{cl}_{\dot x}(\omega) \! = \! \int \langle \dot x(t) \dot x(0)\rangle e^{i\omega t} \mathrm{d}t$ (where $\langle \ldots \rangle$ denotes now averaging over realizations of noise) is related to the classical spectrum of $x(t) $ noise by $S^{cl}_{\dot x}(\omega) \! =\! \omega^2 S^{cl}_{x}(\omega)$, and that $\sin^2\theta \! =\! \tun^2/\Omega_0^2$ and $\cos^2\theta \! =\! (v\tau)^2/\omega_0^2$, we have
\begin{align}
    \Gamma_{\pm,\epsilon}(\tau) & = \frac{1}{4} \frac{\tun^2}{\Omega^2_0(\tau)} S^{cl}_{\epsilon}[\mp\Omega_0(\tau)] \,\, , \label{eq:Gamma_eps} \\
       \Gamma_{\pm,t}(\tau) & = \frac{1}{4} \frac{(v\tau)^2}{\Omega^2_0(\tau)} S^{cl}_{t}[\mp\Omega_0(\tau)] \,\, . \label{eq:Gamma_t}
\end{align}
In these equations the $\mp \Omega_0$ arguments can be, of course, replaced by $|\Omega_0|$, as the classical spectra are symmetric in frequency.
In Appendix \ref{app:stationary_phase} we give an alternative derivation of these results (in the spirit of methods used previously in \cite{Krzywda_PRB20,Malla_PRB17}). We also show there that the AME calculation using these rates agrees very well with direct averaging of evolution due to $\hat{H}(\tau)$ averaged over realizations of classical noise with experimentally relevant parameters (discussed below in this Section). In this way we check the applicability of AME to the system of interest in this paper in the classical noise/high temperature regime.

Eqs.~\eqref{eq:Gamma_eps} and \eqref{eq:Gamma_t} connect the rates as given $\tau$ with (classical) spectra of appropriate noise at $\pm \Omega_0$ frequencies. Extension of AME to regime of lower temperatures/higher $\Omega_0$ amounts then to replacing the classical spectra, $S^{cl}(\pm\Omega_0)$, by their quantum counterparts, $S^{Q}(\pm \Omega_0)$.  Let us now discuss the classical and quantum regimes for the two  charge noise spectra relevant for semiconductor quantum dots in GHz range ($\tun \sim \! 10$ $\mu$eV) of energies.

First we consider electric fluctuations from electron gas in metallic gates, the Johnson-Nyquist noise of general form \cite{huangSpinRelaxationDue2014,Clerk_RMP10,Weiss}:
\begin{equation}
\label{eq:joh}
S^{Q,(J)}_\epsilon(\omega) =  \frac{\Re{Z}}{R_q} \frac{\omega}{1 - e^{-\beta \omega}} 
\end{equation}
where  $R_q$ is the inverse of conductance quantum $R_q =\pi/e^2 =13$ $\text{k} \Omega$ and $Z$ is the impedance of a noise source, which we model here as an ideal resistor (R) of the impedance given by typical for microwaves $Z_R = R = 50\Omega$. The temperature-dependent part of Eq.~\eqref{eq:joh} reduces to Bose-Einstein distribution $n(\omega) = 1/(e^{\beta \omega}-1)$ for negative frequencies $\omega<0$ (absorption) and $n(\omega) + 1$ for $\omega>0$ (stimulated and spontaneous emission).  In the $\gtrsim $ GHz frequency range relevant here, Johnson noise from a lossy transmission line discussed in  \cite{Hollman_PRAPL20} for Si/SiGe quantum dot, gives at most an order of magnitude larger noise power.

Next, we consider $1/f$-type fluctuations of electric field due to two-level fluctuators (TLFs) localized in the insulating regions of the nanostructure \cite{Paladino_RMP14}. We focus first on noise in detuning, as there are numerous measurements of spectrum of this noise in DQDs.
Due to very high spectral weight at low frequencies such a $1/f$ noise dominates the dephasing of qubits the energy splitting of which depends on electric fields \cite{Paladino_RMP14,Szankowski2017}. Here, however, we focus on high (GHz range) positive and negative frequency behavior of $S^{Q,(1/f)}(\omega)$ that is of $1/|\omega|^{\alpha}$ character at very low frequencies. 
The behavior of quantum noise caused by an ensemble of TLFs at such frequencies depends on microscopic details of these fluctuators and the distribution of their parameters, see \cite{youPositiveNegativefrequencyNoise2021} and references therein.

Here, as in \cite{yangAchievingHighfidelitySinglequbit2019}, where Si/SiGe charge qubit in  a DQD was considered, we take $\alpha \!=\! 1$ with noise amplitude directly extrapolated from the low-frequency regime, i.e.~for positive-frequency quantum spectrum we have  
\begin{equation}
\label{eq:1f}
S^{Q,(1/f)}_\epsilon(\omega>0) = s_{1}(T)\frac{\omega_1 }{\omega},
\end{equation} 
where $\omega_1 = 2\pi$/s and $s_{1}(T) =S^{cl}(\omega_1)$ is a commonly reported classical spectral density at $f = 1$Hz, which at electron temperature of $T = 100$mK in typical Si/SiGe device is given by $s_{1}(100\text{mK})\approx (0.3-2)^2\,\mu$eV$^2$/Hz \cite{Mi_PRB18, StruckNPJQ2020,Freeman_APL16,connorsLowfrequencyChargeNoise2019,Petit_PRL18}. 
As $s_1(T) \propto T$ scaling was observed in experiments on quantum dots \cite{Freeman_APL16,Petit_PRL18,connorsLowfrequencyChargeNoise2019}, we assume here $s_1(T) \! =\! s_1(100\text{mK}) \tfrac{T}{100\text{mK}}$. The negative-frequency quantum spectrum follows from Eq.~\eqref{eq:1f} using the detailed balance condition. It is commonly believed that charge disorder in Si-MOS should have larger amplitude, for example $s_1(100$mK$)\approx 10 \mu$eV$^2$/Hz was measured in \cite{kimLowdisorderMetaloxidesiliconDouble2019} at $T =300$mK. However, following \cite{kranzExploitingSingleCrystal2020a} and references therein, we assume that the $1/f$ noise amplitude in Si-MOS can be made comparable or even smaller then Si/SiGe \cite{Freeman_APL16}.

Let us stress that the character of noise generated by an ensemble of TLFs  above $\sim$ MHz frequency is not universal, as its amplitude and exponent varies between the DQDs materials and devices. In particular, recent measurements of charge noise in Si/SiGe \cite{connorsChargenoiseSpectroscopySi2021} and Si-MOS \cite{jockSiliconSinglettripletQubit2021} showed $1/f$ and $1/f^{0.7}$ scaling up to $100$MHz and $1$MHz respectively, which contrasted with few orders of magnitude weaker amplitude of charge noise at MHz frequencies in  some of  GaAs singlet-triplet  qubits \cite{dialChargeNoiseSpectroscopy2013,cerfontaineClosedloopControlGaAsbased2020a}. Additionally, in neither experiment a linear scaling of spectral density with temperature was seen at highest frequencies, and in particular the Si/SiGe case showed only weak dependence on the temperature, confirmed also elsewhere for SiMOS \cite{petitUniversalQuantumLogic2020,kranzExploitingSingleCrystal2020a}, which stood in contrast to GaAs device, where $S(\omega) \propto T^2$ and the spectrum became flat, i.e $\alpha \to 0$ as $T$ was increased \cite{dialChargeNoiseSpectroscopy2013}.
A recent theoretical study \cite{youPositiveNegativefrequencyNoise2021} of qubit relaxation caused by interaction with an ensemble of TLFs coupled to thermal bath (which create $1/f$ noise at low frequencies) showed that at high positive  frequencies (between MHz and GHz, depending on temperature), a crossover first to  $S^{Q}(\omega) \propto 1/\omega^2$, and then to a flat or Ohmic spectrum (depending on details of distribution of energy splitting of the TLFs) occurs.  
One can thus expect that in measurement of high-frequency quantum noise, it is difficult to distinguish the noise caused by TLFs from other sources of electric field fluctuations, as a flat spectrum has been observed already at MHz frequency in SiMOS QD spectroscopy \cite{chanAssessmentSiliconQuantum2018}. Let us note that one of the models of distribution of energies of TLFs considered in \cite{youPositiveNegativefrequencyNoise2021} led to $S^{Q}(\omega >0) \! \propto \! T$ at high frequencies. 
In light of the above discussion, we use the above model to estimate the relevance of the  tail of 1/f noise in silicon-based devices and set $s_1(0.1\text{K}) \! =\!  1^2\mu$eV$^2$/Hz. We will use the same spectrum for GaAs, probably overestimating the noise in this case, but below we will show that for GaAs quantum dots the influence of electron-phonon coupling dominates over that of charge noise having even such a large amplitude. 

For the charge noise in tunnel coupling, we assume that it is uncorrelated with the noise in detuning, as the two are caused by TLFs from distinct spatial regions.
We parametrize the ratio of rms of fluctuations of the noise in $\tun$ and $\epsilon$ by $\eta = \sqrt{S_{t}(\omega)/S_{\epsilon}(\omega)} \! \approx \! 0.1 $,
with its value motivated by semiconductor quantum dots experiments \cite{nakajimaCoherentTransferElectron2018,Shi_PRB13,dialChargeNoiseSpectroscopy2013, connorsChargenoiseSpectroscopySi2021}, and typical values of level arm used to control the electronic gates during shuttling \cite{Mills_NC19}. We conclude this section by giving the explicit forms of longitudinal and transverse contributions to relaxation rates due to charge noise
 \begin{align}
     \Gamma_{-,\epsilon}(\tau) &= \frac{1}{4} \left(\frac{t_c}{\Omega(\tau)}\right)^2 S^Q_\epsilon[\Omega_{0}(\tau)] \,\, , \\
     \Gamma_{-,t}(\tau) &= \frac{\eta^2}{4} \left(\frac{v\tau}{\Omega(\tau)}\right)^2 S^Q_\epsilon[\Omega_{0}(\tau)] \,\, ,
\end{align}
which are applicable for both $1/f$ and Johnson noise. The corresponding excitation rates are obtained via detailed balance condition $\Gamma_+(\Omega) = \Gamma_- (\Omega) e^{-\beta \Omega}$.

\subsection{Electron-phonon interaction}
\label{sec:phonon_bath}
In semiconductors, another mechanism responsible for transitions between the  $\ket{\pm,\theta}$  states is associated with energy exchange between the electron and lattice vibrations. Phonons are assumed to be in thermal equilibrium, with their free Hamiltonian given by $H_{\text{ph}} = \sum_{\kb,\lambda} \omega_{\kb,\lambda} b^\dagger_{\kb,\lambda}b_{\kb,\lambda} $, where $\lambda$ and $\mathbf{k}$ represents phonon polarizations and wavevector respectively. The electron-phonon interaction is given by \cite{Raith_PRL12}:
\begin{align}
\label{eq:el-ph}
\mathcal{H}_\text{el-ph} &= \sum_{j,\kb,\lambda = L,T}\sqrt{\frac{|\mathbf{k}|}{2\varrho c_\lambda V}}\, v^{(j)}_{\kb,\lambda}  \big(\hat b_{\kb,\lambda} + \hat b_{-\kb,\lambda}^\dagger \big)e^{i\kb \mathbf{r}}  \,\, ,
\end{align}
in which $\varrho$ denotes crystal density, $V$ the crystal volume, and $c_\lambda$ is the speed of $\lambda$-polarized phonons. The coupling  $v_{\kb,\lambda}^{(j)}$ stands for piezoelectric ($j=p$) and deformation potential ($j=d$), evaluated for transverse ($\lambda = T$) and longitudinal ($\lambda = L$) polarizations of phonons:
\begin{equation}
\label{eq:coupling_fun}
 v^{(\text{p})}_{\kb,\lambda} = \frac{\chi_p}{k},\quad v^{(\text{d})}_{\kb,L}=  \Xi_d + \Xi_u \left(\frac{k_z}{k}\right)^2,\quad v^{(\text{d})}_{\kb,T} =-\Xi_u \frac{k_{xy}k_z}{k^2},
\end{equation}
where $\chi_p$ is piezoelectric constant, while $\Xi_d$, $\Xi_u$ are dilatation and shear deformation potentials respectively.
In GaAs and Si the coupling to phonons takes a very different form, namely Si lacks the dominant in the GaAs piezoelectric coupling $\chi_p^\text{SiGe}=0$ \cite{yuFundamentalsSemiconductors2010}, while the opposite is true for shear deformation potential since $\Xi_u^{\text{GaAs}}=0$. 

 We evaluate the matrix elements of interaction from Eq.~\eqref{eq:el-ph} in the two-dimensional space spanned by $\ket{L/R}$ states (see Appendix \ref{app:phonons} for details), obtain the $\hat{V}_t \hat\sigma_x+\hat{V}_\epsilon\hat\sigma_z$ form of coupling discussed in Sec.~\ref{sec:rates_general}, and arrive at quantum spectra associated with longitudinal ($\hat{V}_\epsilon$) and transverse ($\hat{V}_t$) couplings to phonons:
   \begin{equation}
  \label{eq:spectrum_ph}
     S_{\epsilon/t}^{\text{(ph)}}[\Omega(\tau)] =  4\pi\sum_{j, \kb, \lambda} \frac{ k| v_{\kb,\lambda}^{(j)}|^2}{\varrho c_\lambda V} |\mathcal{M}_{\epsilon/t}(\mathbf{k})|^2 \frac{\delta(\Omega(\tau) - \omega_{\kb,\lambda})}{1-e^{-\beta \omega_{\kb,\lambda} }},
 \end{equation}
 where the matrix elements read $\mathcal{M}_{\epsilon}(\mathbf{k}) = \bra{L}e^{i\mathbf{k}\mathbf{r}}\ket{L} -\bra{R}e^{i\mathbf{k}\mathbf{r}}\ket{R}$, and $\mathcal{M}_{t}(\mathbf{k}) = \bra{L}e^{i\mathbf{k}\mathbf{r}}\ket{R} +\bra{R}e^{i\mathbf{k}\mathbf{r}}\ket{L}$, while the temperature-dependent term reduces to Bose-Einstein distribution $n(\Omega)$ for $\Omega<0$ (absorption) and to $n(\omega)+1$ for $\Omega >0$ (emission). The transition rates are given by
 \begin{align}
 \Gamma_{\pm,\epsilon}^{(\text{ph})}(\tau) &= \frac{1}{4}\sin^2\theta\, S_\epsilon^{(\text{ph})}[\mp \Omega(\tau)] \label{eq:Gph_epsilon}   \\ 
 \,\Gamma_{\pm,t}^{(\text{ph})}(\tau) &= \frac{1}{4}\cos^2\theta\, S_t^{(\text{ph})}[\mp \Omega(\tau)] \,\, . \label{eq:Gph_t}  
 \end{align}

For further calculation we need to specify a model of $\bra{\mathbf{r}}L_0/R_0\rangle$ wavefunctions localized in the uncoupled dots. We assume that they are separable and Gaussian: 
\begin{equation}
\label{eq:rL0R0}
    \bra{\mathbf r}\ket{L_0/R_0} = \frac{1}{(\pi^3 r_{xy}^4 r_{z}^2)^{\frac{1}{4}}} \exp{-\frac{(x\mp \tfrac{\Delta x}{2})^2+y^2}{2r_{xy}^2}-\frac{z^2}{2r_{z}^2}},
\end{equation}
where the full width at half maximum (FWHM) of electron wavefunction, which defines \textit{dots diameter}, is given by $2r_{xy}$ in planar, and $2r_z$ in the growth direction of structure, while the $\Delta x$ gives the distance between the dots. Next we use Hund-Mulliken approximation \cite{Burkard_PRB99,Li_PRB10}, to generate a set of orthogonal states in DQDs system that fulfill $\bra{L}\ket{R}=0$, which can be done by setting:
\begin{equation}
\ket{L/R} = \ket{L_0/R_0} - g\ket{R_0/L_0},
\end{equation}
where $g \! \approx \! \frac{1}{2} \bra{L_0}\ket{R_0} = \tfrac{1}{2}e^{-\Delta x^2/4r_{xy}^2}\ll 1$.

As the energy quanta exchanged between the electron and the lattice are $< \! 1$ meV, we take into acocunt only the acoustic phonons with  $\omega_{\kb,\lambda} = c_\lambda |\kb| $.
The relaxation rates due to electronphonon interaction are then given by:
\begin{align}
\Gamma_{-,\epsilon}^{\text{(ph)}}(\Omega) &= \sum_{\lambda,j } \frac{\Omega
\, t_c^2}{8\pi^2\varrho c_\lambda^5} \int\text{d}\Omega_\kb \big|v_{\kb_\lambda,\lambda}^{(j)}\big|^2 |F_{\epsilon}(\kb_\lambda)|^2 [n(\Omega)+1] \label{eq:gam_ph_e} \\
\Gamma_{-,t}^{\text{(ph)}}(\Omega) &= \sum_{\lambda,j } \frac{\Omega 
\, \epsilon^2}{8\pi^2\varrho c_\lambda^5} \abs{\bra{L_0}\ket{R_0}}^2 \int \text{d}\Omega_\kb\big|v_{\kb_\lambda,\lambda}^{(j)}\big|^2 |F_{t}(\kb_\lambda)|^2  \nonumber \\
& [n(\Omega)+1] \,\, , \label{eq:gam_ph_t}
\end{align}
where the integration over solid angle of resonant wavevector $\kb_\lambda$, with length $k_\lambda = \Omega/c_\lambda$  was denoted by $d\Omega_\kb = d\vartheta_{\kb}d\varphi_{\kb}\sin\vartheta_{\kb}$, while the form factors read:
\begin{align}
\label{eq:Fxz}
|F_t(\kb)|^2 &= \exp(-\frac{k_{xy}^2 r_{xy}^2 + k_z^2 r_z^2}{2})\, \big(1-\cos(\tfrac{k_x\Delta x}{2})\big)^2\nonumber \\ 
|F_\epsilon(\kb)|^2 &=\exp(-\frac{k_{xy}^2 r_{xy}^2 + k_z^2 r_z^2}{2})\,\sin^2(\tfrac{k_x\Delta x}{2}).
\end{align}
The common term $\exp(-(k_{xy}^2 r_{xy}^2 + k_z^2 r_z^2)/2)$ is the Fourier transform of the electron wavefunction, while the main difference between the longitudinal and transverse relaxation is the overlap of bare dots wavefunctions, $\abs{\bra{L_0}\ket{R_0}}^2 = e^{-\Delta x^2/2r_{xy}^2} \ll 1$ which makes the transverse relaxation mechanism orders of magnitude weaker, i.e. $\Gamma_{-,t}^{(\text{ph})} \ll \Gamma_{-,\epsilon}^{(\text{ph})}$, unless detuning is so large that $\theta$ is close enough to $\pi$ for $\sin^2\theta$ term in Eq.~\eqref{eq:Gph_epsilon} suppresses $\Gamma_{-,\epsilon}^{(\text{ph})}$ to the degree that is becomes smaller than $\Gamma_{-,t}^{(\text{ph})}$.

\begin{widetext}

\label{app:parameters}
 \begin{table}[h!]
 \centering
\begin{tabular}{|l|c|c|}
\hline	
Quantity & Symbol & Values \\ \hline \hline
Tunnel coupling & $t_c$ & $5-60\mu$eV \\ \hline
Effective electron temperature & $T$ & $50-500$mK \\ \hline
Detuning sweep rate & $v$ & $1-3000\mu$eV/ns \\ \hline
Initial detuning& $\epsilon_i$ & $-500\mu$eV \\ \hline
Final detuning& $\epsilon_f$ & $500\mu$eV \\ \hline
Time of detuning sweep & $\epsilon_f-\epsilon_i= \Delta \epsilon/v$ & $0.3-1000$ns \\ \hline
Transverse/longitudinal noise ratio & $\sqrt{S_t(\omega)/S_\epsilon(\omega)}$ & $0.1$ \\ \hline
Resistance of noisy resistor (Johnson noise) & $Z_R$ & $ 50\Omega$ \\ \hline 
1/f noise amplitude at $T=0.1$K& $s_{1}(0.1$K$)$ & $1\mu\text{eV}^2/\text{Hz}$ \\ \hline
Dots separation & $\Delta x$ & $50$nm (GaAs), $100$nm (SiGe), $150$nm (Si-MOS) \\ \hline
Spread of electron wavefunction in XY plane  & $r_{xy}$ & $40$nm (GaAs), $20$nm (Si/SiGe, Si-MOS) \\ \hline
Width of quantum well & $2r_{z}$ & $20$nm (GaAs),$5$nm (Si/SiGe, Si-MOS) \\ \hline
\end{tabular}
\caption{Parameters used in the paper.}
\label{tab:parameters_sim}
\end{table} 

\end{widetext}

\begin{figure}[tbh!]
\includegraphics[width=\columnwidth]{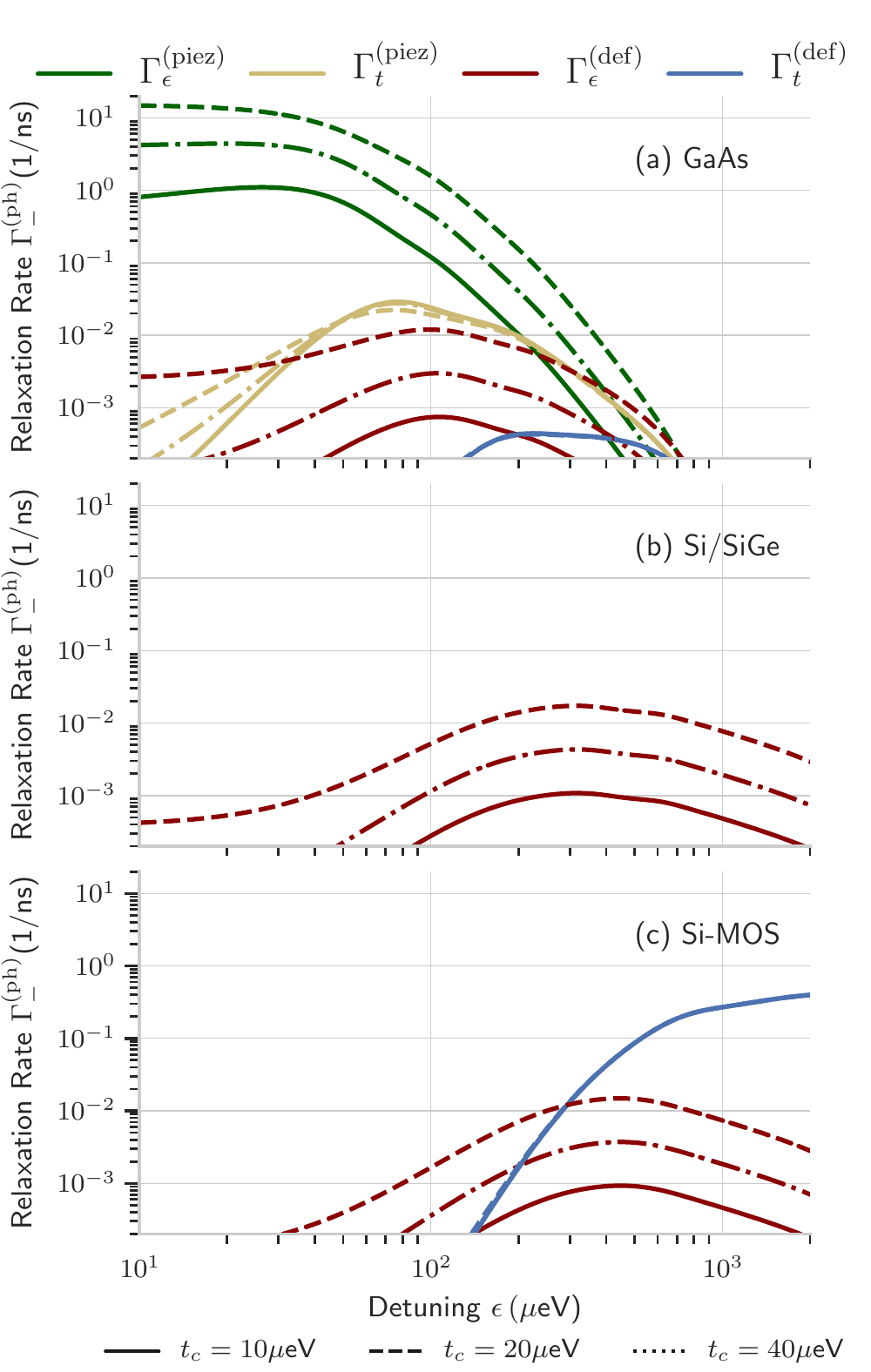}
\caption{Phonon relaxation rates in (a) GaAs, (b) Si/SiGe, (c) Si-MOS double quantum dot devices as a function of dots detuning at fixed tunnel couplings: $t_c = 10\mu$eV (solid line), $20\mu$eV (dashed-dotted) and $40\mu$eV (dashed). Contributions from different phonon mechanisms are shown with a distinct color: piezoelectric longitudinal coupling $\Gamma_\epsilon^{(\text{piez})}$ (green), piezoelectric transverse coupling $\Gamma_t^{(\text{piez})}$ (yellow), deformation longitudinal coupling $\Gamma_\epsilon^{(\text{def})}$ (red) and deformation transverse coupling $\Gamma_t^{(\text{def})}$ (blue). Longitudinal phonons couples orbital-like states in vincinity of avoided crossing, while tranverse phonons couples dot-like states in detuned regime. Each panel represents different device with parameters given in Tab.~\ref{tab:parameters_sim}.}
\label{fig:phonons}
\end{figure}

As the size of quantum dot in planar direction $r_{xy}$ is larger then size in the $z$ direction, $2r_z \ll r_{xy}$, its value can be extracted from splitting between the ground and first excited dot state $\Delta E$, i.e. $r_{xy} = \sqrt{1/m^* \Delta E}$. In Si at $\Delta E = 1$meV an estimate of $r_{xy}\sim 20$nm is consistent with reported values of $r_{xy} \approx 15$nm \cite{Wang_PRB13}, $13$nm \cite{StruckNPJQ2020} in Si/SiGe and $r_{xy} \approx 21$nm \cite{eeninkTunableCouplingIsolation2019a}, $18$nm \cite{kimLowdisorderMetaloxidesiliconDouble2019} in Si-MOS. The GaAs dots are typically larger ($r_{xy} \approx 55$nm \cite{MalinowskiPRL2017}, $21$ nm \cite{SrinivasaPRL2013}), mostly due to smaller effective mass, i.e. $m^*_{\text{Si}}/m^*_{\text{GaAs}} \approx 3$. The typically reported values of $2r_z \approx 20$nm \cite{MalinowskiPRL2017} in GaAs are also larger than those in Si/SiGe, $2r_z \approx 4$nm \cite{WangPRL2013}, $6$nm \cite{StruckNPJQ2020}. We assume here the extent of electron's wavefunction in the $z$ direction in Si-MOS is similar to that in Si/SiGe, and for both we take it as $2r_z = 5$nm. Finally, smaller dots allow for decreasing the distance between the sites from typical for GaAs $\Delta x \sim 150$nm \cite{MalinowskiPRL2017}, $\sim 110$ nm \cite{SrinivasaPRL2013} to Si/SiGe values of $\Delta x \sim 100$nm \cite{lawrieQuantumDotArrays2020a}, to Si-MOS $\Delta x \sim 50$nm \cite{lawrieQuantumDotArrays2020a}. The distances between the dots are correlated with reported values of $t_c$, the largest of which are achieved in SiMOS structures, with examples of $t_c \approx 450\mu$eV and $50\mu$eV for dots separated by $\Delta x \sim 40$nm \cite{yonedaCoherentSpinQubitc} and $\sim100$nm \cite{eeninkTunableCouplingIsolation2019a} respectively.
However, recently $\tun \! \approx \! 40\mu$eV was achieved in Si/SiGe across an array of quantum dots with $r_{xy} \!\sim\! 10$nm and $\Delta x \!\approx\! 70$nm \cite{Mills_NC19}. In GaAs, tunnel coupling of $t_c\! \approx \!20-40\mu$eV was measured  in an array of eight quantum dots with $\Delta x \!\approx \!150$nm  \cite{Volk_NPJQ19}  for array of 8 QDs. Representative parameters for each nanostructure that we will use in subsequent calculations are given in Tab.\ref{tab:parameters_sim}.  

We now evaluate numerical values of relaxation rates from Eqs.~\eqref{eq:gam_ph_e} and\eqref{eq:gam_ph_t} for above-discussed parameters of ``typical'' GaAs, Si/SiGe and Si-MOS double quantum dots given in table \ref{tab:parameters_sim},
In Fig.~\ref{fig:phonons} we plot zero-temperature electron relaxation rate due to scattering with phonons, $\Gamma_-^{(\text{ph})}(\Omega)$, as a function of detuning $\epsilon$ (let us recall that $\Omega = \sqrt{t_c^2 +\epsilon^2}$) for three values of tunnel coupling,  $t_c = 10,20,40\mu$eV. It is clear that the scattering of a single electron in a DQD in  each of considered nanostructures is dominated by a different mechanisms. In polar GaAs, the piezoelectric coupling dominates over the deformation potential one, with the fastest relaxation at low detuning, where the transitions occur between molecular-orbital type states.
The relaxation rate, for the energies below $c/\Delta x \! \approx \! 50\mu eV$ shows oscillatory behaviour due to $|F_\epsilon|^2 \propto \sin^2(k_x\Delta x/2)$ term, see Eq.~\eqref{eq:Fxz}. For larger detunings, when the energy transfer $\Omega \! \approx \! \epsilon$, the relaxation rate decreases as its mostly longitudinal character that makes it $\propto (t_c/\epsilon)^2$, is combined with phonon spectral density $ \propto \epsilon^3$ and piezoelectric coupling $|v^{(\text{piez}})|^2\propto \epsilon^{-2}$, to produce an overall $\Gamma_{-}^{(\text{piez})} \propto (t_c)^2/\epsilon$ scaling in the far detuned regime $\epsilon \gg t_c$, until $\epsilon \! \approx \! 500\mu$eV when phonon bottleneck effects start to become strongly viisble.
On the other hand, in Si/SiGe a weaker deformation potential scattering gives $\Gamma_{\epsilon}^{(\text{def})}$ that first increases with $\epsilon$, and then becomes suppressed by phonon bottleneck effect at large detunings. The relaxation time $1/\Gamma_-$ falls below 100 ns for $\epsilon \! \sim \! 100$ $\mu$eV only for the largest considered $t_c = 40\mu$eV. Finally, in Si-MOS the smaller interdot distance makes the transverse relaxation more efficient. Due to its  $\Gamma_{t,-}^{(\text{def})} \propto \epsilon^3$ scaling up to phonon bottleneck energy of about $1$ meV, it becomes the dominant process at larger detunings. Such a transverse relaxation rate weakly depends on tunnel coupling (note the presence of single blue lines in Fig.~\ref{fig:relax}, and requires overlap between wavefunctions of L/R dots, which is not large enough in the other nanostructures: $\Gamma^{(\text{piez})}_t$ might be relevant in GaAs only at highest detunings, see Fig.~\ref{fig:relax}a, and the transverse process never becomes of similar order of magnitude as the longitudinal one in the considered Si/SiGe structures.

\subsection{Comparison of the transition rates}

\begin{figure}[tbh!]
\includegraphics[width=\columnwidth]{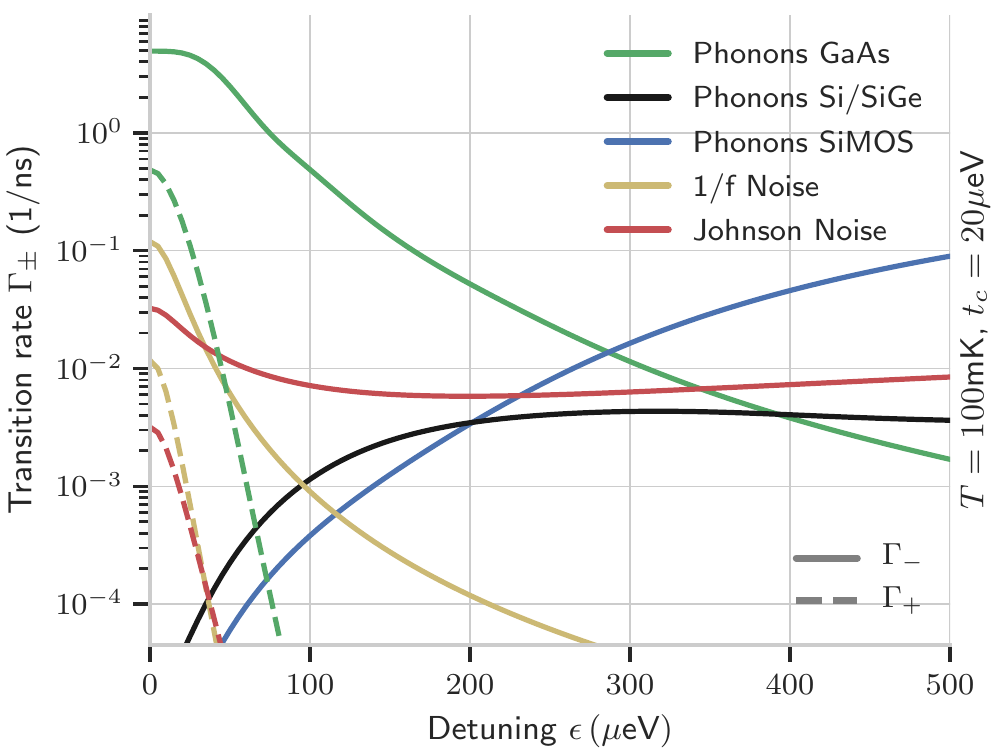}
\caption{Relaxation (solid lines) and excitation (dashed lines) rates as a function of detuning at typical tunnel coupling $t_c = 20\mu$eV and temperature of $T = 100$mK. 
Transition rates due to phonons are drawn using green (GaAs), black (Si/SiGe) and blue (Si-MOS) colors, while transition rates due to common for all nanostructures charge noise is depicted using red (Johnson) and yellow (1/f) colors. Both excitations and relaxations in GaAs are dominated by electron-phonon coupling. In Si the excitations are commonly caused by charge noise (either 1/f or Johnson of similar amplitude), while the relaxation at finite detuning relies on Johnson noise in Si/SiGe and relatively stronger interdot phonons in Si-MOS, where the dots are closer.}
\label{fig:relax}
\end{figure}

\begin{figure}[tbh!]
\includegraphics[width=\columnwidth]{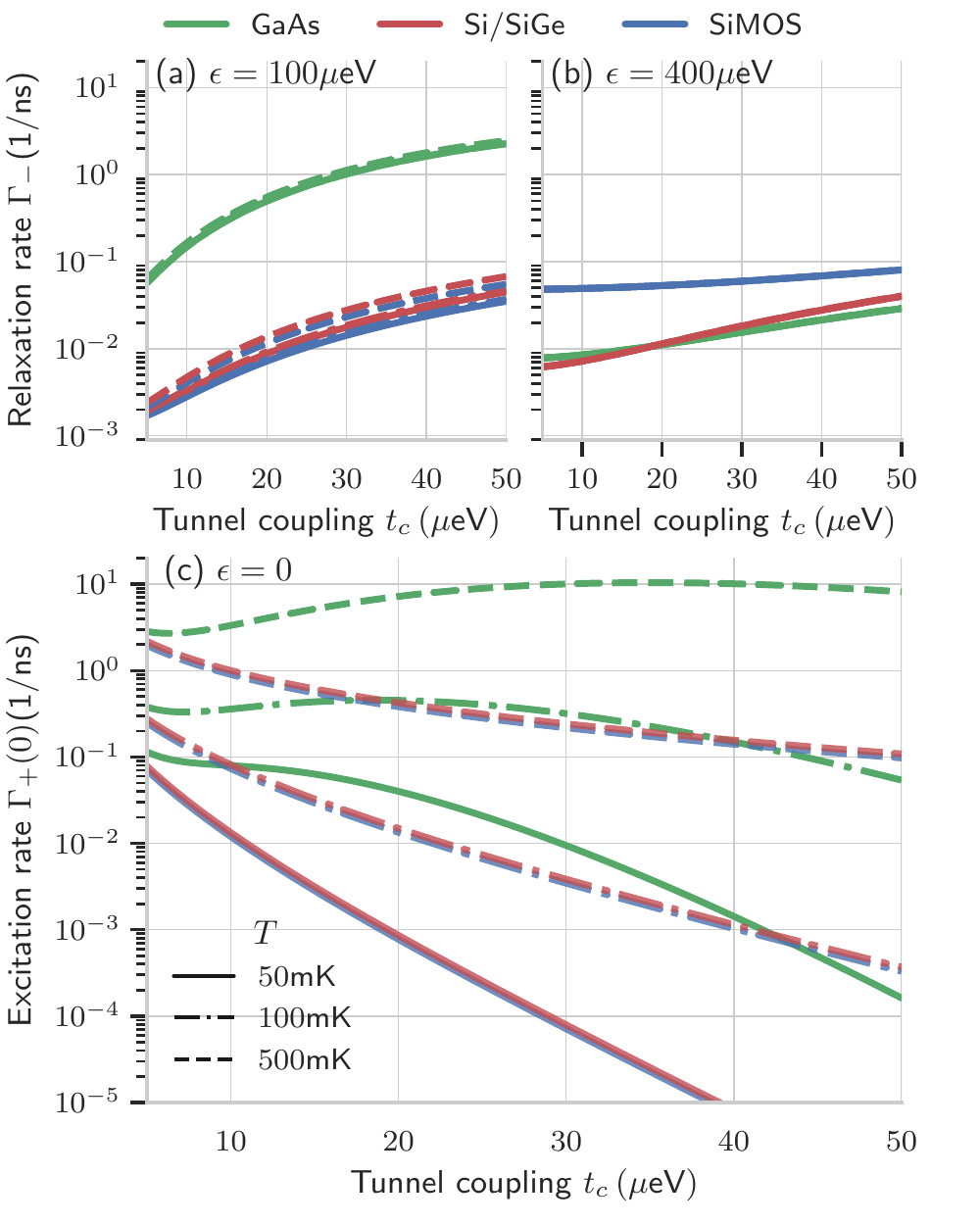}
\caption{Transition rates relevant for electron charge transfer in GaAs (green), Si/SiGe (red) and Si-MOS (blue): the excitation rate at avoided crossing $\Gamma_+(0)$ ($\epsilon=0$) (a), relaxation rates in detuned regime $\Gamma_-(\epsilon=100\mu$)eV (b) and $\Gamma_-(\epsilon=100\mu$)eV (c) as a function of tunnel coupling $t_c$ for selection of temperatures $T = 50,100,500$mK (solid, dashed-dotted, dashed) lines respectively. The excitation rate due to piezoelectric phonons in GaAs in $T>100$mK is the only non-monotonic function of tunnel coupling. Otherwise excitation rates decreases for larger $t_c$ due to exponential factor $\Gamma_+(0) \propto e^{-\beta t_c}$, while relaxation rates increases due to dominant role of longitudinal mechanisms $\Gamma_-(\Omega)\propto t_c^2/\Omega^2$. Since $\Omega = \sqrt{t_c^2 + \epsilon^2}$ increase is stronger at lower $\epsilon$. The only discrepancy between Si/SiGe and Si-MOS is visible in the relaxation rate at far detuned regime ($\epsilon = 400\mu$eV) due to presence of interdot phonons in the latter.}
\label{fig:relax_exc}
\end{figure} 

Let us know use the results of the previous Sections and compare the relative importance of various types of environments on the discussed DQD structures. In Fig.~\ref{fig:relax} we plot the relaxation rate $\Gamma_-[\Omega(\epsilon)]$ (the excitation rate $\Gamma_+(\Omega)\! =\! \Gamma_{-}(\Omega)e^{-\beta\Omega}$) with solid (dashed) lines as functions of detuning for all the considered mechanisms using the above-discussed representative parameters for GaAs, Si/SiGe, and SiMOS structures, and  temperature $T = 100$mK and tunnel coupling $t_c = 20\mu$eV. As expected from discussion in Sec.~\ref{sec:approximate}, the excitation rates are the largest at the anticrossing, and they become suppressed exponentially with $\epsilon$ increasing above $\tun$. In that regime the relaxation overwhelmingly dominates over excitation, but the dynamics of the electron will depend on the value of total $\Gamma_{-}$: the electron tranfer error will depend on the ratio of timescale of environment-assisted inelastic tunneling between the dots in the far-detuned regime, $1/\Gamma_-$, and the duration of the detuning sweep. Note that for $\tun\! =\! 20$ $\mu$eV the requirement of $Q_{LZ} \! < \! 10^{-4}$ means $v\! <\! 200\mu$eV/ns, so the total time of detuning sweep over a meV range is $5$ns. This will give a ballpark estimate what timescales we should compare $1/\Gamma_-$ to.

In GaAs the coupling to phonons (the green line in Fig.~\ref{fig:relax} dominates the relaxation, with influence of Johnson noise possibly becoming dominant at highest considered detunings. As discussed above, $\Gamma_{-}^{(\text{piez})} \propto (t_c)^2/\epsilon$ for most of the considered range of $\epsilon$, so for the healing of excitation to be significant the time spent at moderate detunings, up to about 200 $\mu$eV (see Fig.~\ref{fig:relax}), has to be larger than average relaxation time in this range, $1/\Gamma_- \! \sim \! 1-10$ ns.

The situation is more complex in Si nanostructures. For parameters of Si/SiGe DQDs it is the Johnson noise - red line in Fig.~\ref{fig:relax} - that dominates (more visibly at lower $\epsilon)$) over the relaxation due to deformation potential coupling to phonons (the black line in Fig.~\ref{fig:relax}). The detuning dependence of this process is rather weak. When $t_c/k_{\text{B}}T \gg 1$ (in Fig.~\ref{fig:relax} we have $t_c/k_{\text{B}}T \approx 2.3$), the Johnson noise from $50\Omega$ resistor gives $\Gamma^{(\text{J})}_{-,\epsilon} \propto t_c^2/\epsilon$ for stronger longitudinal process, and and $\Gamma^{(\text{J})}_{-,t} \propto \epsilon$ in case of weaker transverse one. For their assumed ratio, the relaxation the rates become equal at $\epsilon_J = 10t_c$, which means $\Gamma^{(J)}_-$ slowly decreases as $\epsilon^{-1}$ up to $\epsilon= 400\mu$eV, and then it starts to slowly increase with $\propto \epsilon$. The relaxation time for the assumed amplitude of Johnson noise is $\sim \! 100$ ns in the relevant detuning range. 

Finally, for SiMOS the smaller interdot distance assumed for this architecture makes $\Gamma_{t,-}^{(\text{piez})} \propto \epsilon^3$ the dominant relaxation process at large detunings: as show in Fig.~\ref{fig:relax} this relaxation channel dominates over the one due to Johnson noise for $\epsilon \! \gtrsim \! 200$ $\mu$eV. The relaxation times at large  detunings approach $\sim \! 10$ ns, so phonon-assisted interdot tunneling might be an efficient mechanism of healing of charge noise-induced excitation that occurred close to the anticrossing in SiMOS.

The other mechanisms only weakly contribute to relaxation, as longitudinal 1/f noise relaxation rate is strongly attenuated with increasing detuning, as $\Gamma_{-,\epsilon}^{(1/f)} \propto t_c^2/\epsilon^3$ at $\epsilon>t_c$, while small overall strength and weak detuning dependence of longitudinal phonon processes in Si/SiGe, $\Gamma_{-.\epsilon}^{(\text{def})} \propto t_c^2 \epsilon$, produces relaxation times above $100$ ns only approaching the order of magnitude of contribution of Johnson noise around $\epsilon = 200\mu$eV.

Let us now discuss the tunnel coupling and temperature dependence of the total $\Gamma_+$ rate at $\epsilon \! =\! 0$, and of the total $\Gamma_-$ rate at moderate and high detunigs, $\epsilon\! =\! 100$ and $400$ $\mu$eV, respectively. The relaxation rates at moderate detuning have a common dependence on $\tun$ inherited from the tunneling dependence of the dominant there longitudinal process, i.e. $\Gamma_- \propto t_c^2$. This is not the case at larger detuning, where transverse processes that are weakly dependent on $\tun$ can dominate. 
Similarly for considered here $\beta t_c \gg 1$, temperature dependence of relaxation is very weak. We illustrate both statements in Fig.~\ref{fig:relax_exc}a and \ref{fig:relax_exc}b where we plot relaxation rates at $\epsilon = 100,400\mu$eV as a function of tunnel coupling. As it can be seen difference between Si/SiGe and SiMOS is visible at large detuning where for small tunnel couplings interdot phonons provide order of magnitude faster relaxation rate in the latter.

In Fig.~\ref{fig:relax_exc}c we illustrate the relevant excitation rate $\Gamma_+(0)$, computed at the avoided-crossing at $T = 50,100,500$mK. 
In GaAs the only relevant mechanism is the coupling between orbital states provided by the phonons, which has a strong scaling with tunnel coupling $\Gamma_+^{(\text{piez)}}(0) \propto t_c^3 e^{-\beta t_c}$ as long as $t_c \ll c/\Delta x \sim 50\mu$eV, wher $t_c^3$ dependence is provided by the piezoelectric coupling ($t_c$) and the resonance term $\sin^2(k_x \Delta x/2)$ ($t_c^2$). As a result, at smaller temperatures, excitation rate in GaAs shows a non-monotonic behaviour as a function of tunnel coupling. In Si, the excitation at $\epsilon \ll t_c$ is caused only by charge noise, and hence for amplitude of this noise used here for both Si/SiGe and SiMOS, it results in the same rate, in which contributions from $\Gamma_+^{(1/f)}(0) \propto e^{-\beta t_c}/t_c$ and $\Gamma_+^{(\text{J})}(0) \propto t_c e^{-\beta t_c}$ are combined. The latter becomes more relevant at larger $\tun$, for which however the overall charge noise is attenuated due to exponential factor, as it can be seen in Fig.~\ref{fig:relax_exc}c by a decrease of excitation rate in Si.


\section{Probability of leaving the electron behind} \label{sec:results}
Let us use now the above-derived transition rates to calculate the central quantity of this paper -- occupation of higher energy state after detuning sweep $Q$, i.e. the probability of leaving the electron in the initial dot.

\begin{figure*}[t!]
\includegraphics[width=\textwidth]{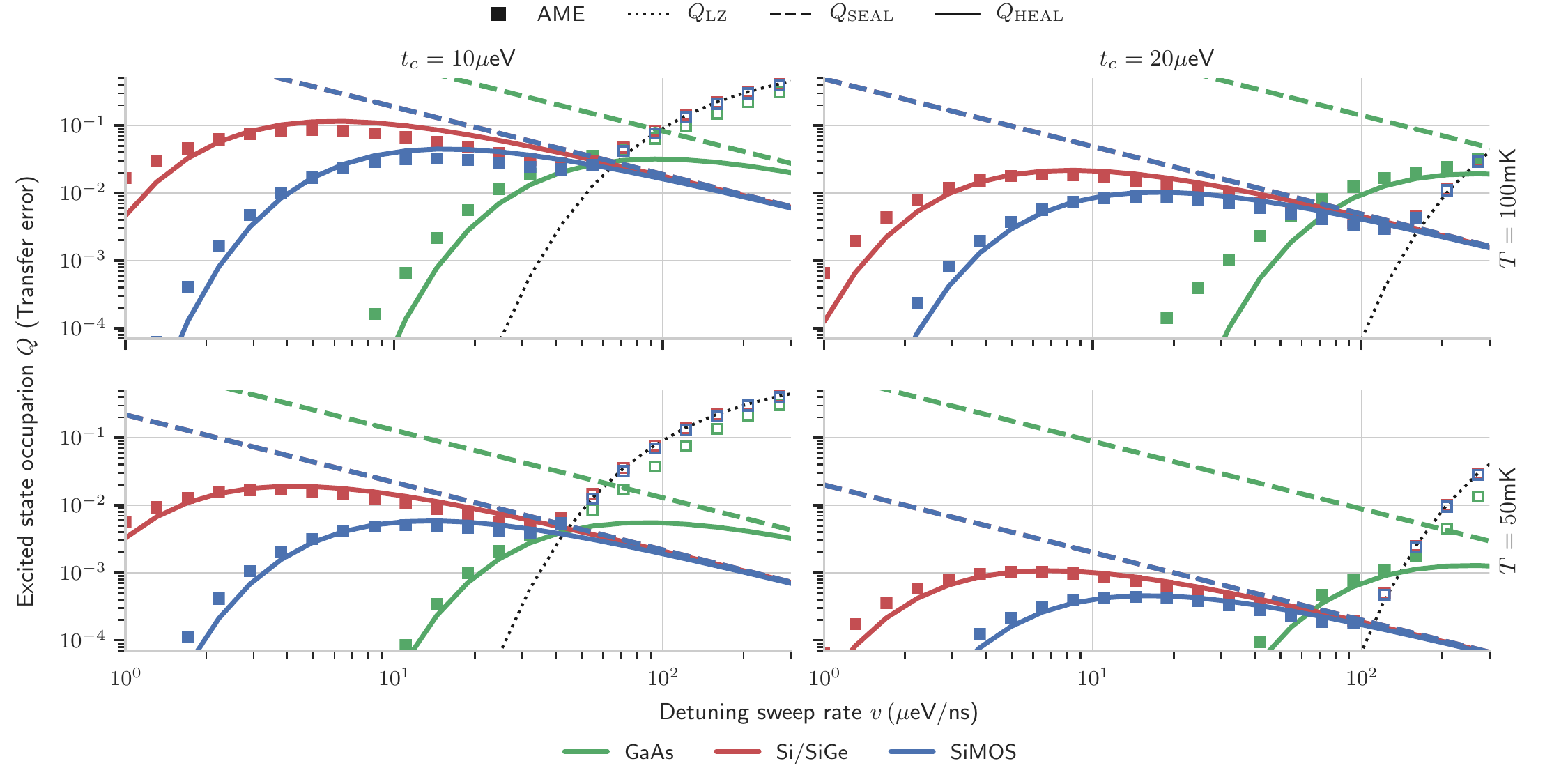}
\caption{Probability of occupying excited state $Q$, i.e.~leaving the electron in the initial dot after detuning sweep, as a function of sweep rate $v$ for fixed tunnel coupling and temperature in two semiconductor DQD devices: GaAs A (Green) and SiGe A for Johnson noise originating from ideal resistor (red) and from transmission line (pink). In the four panels we show combinations of tunnel couplings $t_c = 10,20\mu$eV (columns) and temperatures $T = 50,100$mK (rows). Squares correspond to numerical solution of Adiabatic Master Equation, where we have used filled (hollow) squares to denote adiabatic (non-adiabatic) regime. Dashed line corresponds to Single Excitation Approximation Limit $Q_{\text{SEAL}}$, see Eq.~\eqref{eq:seal}, while the solid line is the Healed Excitation Approximation Limit, $Q_\text{HEAL}$, see Eq.~\eqref{eq:q1chi}. Dotted black line shows the Landau-Zener result $Q_{\text{LZ}} = \exp(-\pi t_c^2/2v)$. Remaining parameters are given in Tab.~\ref{tab:parameters_sim} in the Appendix.~\ref{app:parameters}. }
\label{fig:SiGe_GaAs}.
\end{figure*}

We assume the relevant part of detuning sweep starts and terminates at $\epsilon_f = \pm 500\mu$eV, since at $\epsilon \geq10 t_c$ the dots become uncoupled, i.e.~the approximation of constant $\tun$ breaks down \cite{nakajimaCoherentTransferElectron2018, Medford_PRL13}, and the detuning sweep rate used in an experiment can be increased \cite{Mills_NC19}. In Fig.~\ref{fig:SiGe_GaAs} we compare the results of a numerical solution of Adiabatic Master Equation (AME) from Eq.~\eqref{eq:rate_H0}, depicted as squares, against the approximation of single excitation at avoided crossing without relaxation process, $Q_{\text{SEAL}}$ from Eq.~\eqref{eq:q1}, shown as dashed lines, and the approximation of an excitation followed by  relaxation processes only, $Q_{\text{HEAL}}$ from Eq.~\eqref{eq:q1chi}, shown as solid lines. The dotted line is the Landau-Zener formula $Q_{\text{LZ}}$ from Eq.~\eqref{eq:QLZ}. In the four panels we show results for combinations of tunnel coupling and temperatures: $t_c = 10,20\mu$eV (columns), $T = 50,100m$K (rows). With hollow squares we mark the AME results in the region where $Q_\text{SEAL}$ and $Q_{\text{HEAL}}$ are no longer an upper and a lower bound on $Q$, as probability of Landau-Zener transition dominates. We stress that in this region the applicability of AME in secular approximation used here is limited \cite{yamaguchiMarkovianQuantumMaster2017, arceciDissipativeLandauZenerProblem2017}, however a correction to the L-Z formula   computed using different methods correction is expected to be small for prediminantly longitudinal relaxation $\Gamma_\epsilon \gg\Gamma_t$ \cite{pokrovskyFastQuantumNoise2007,nalbachAdiabaticMarkovianBathDynamics2014,javanbakhtDissipativeLandauZenerQuantum2015}. 

Let us now concentrate on the region in which $Q$ is dominated by effects related to interaction with the thermal environment, where the AME results are plotted as filled squares. For both Si-based devices in a region of moderate $v$ we observe that $Q \propto 1/v$. This suggests that the value of $Q$ follows from a finite excitation probability in a limited range of detunings (near the anticrossing), and the occupation of the excited state grows with increased time spent in this region. In agreement with this picture, $Q \approx Q_{\text{SEAL}}$ (dashed line), and the electron undergoes a single transition from ground to excited state in vicinity of the anticrossing. We note such a transition from ground to excited state at $\epsilon\ll t_c$ in SI-based devices is caused solely by charge noise. As the sweep rate is decreased, an increasing time spent during the electron trasnfer in  the far-detuned regime, $\epsilon\! \gg \! \tun$, allows for a significant recovery of ground state occupation by the relaxation mechanism, which is reflected by a deviation from a SEAL approximation and $Q\sim Q_{\text{HEAL}}$ (solid lines) for smallest sweep rates. The healing effect is stronger for the SiMOS device, due to effective phonon relaxation between the dot-like eigenstates at large detunings. The agreement between the result of evaluation of AME, and the approximation is more visible at lower $T$ (higher $t_c$), since this agreement is expected to improve as $t_c/k_\text{B}T \gg 1 $. In contrast, in GaAs device, $Q$ (plotted in green color) decreases monotonically as the sweep rate $v$ gets smaller. This is a consequence of a much stronger coupling to environment (specifically piezoelectric coupling to phonons), which on one hand increases probability of transition from ground to excited state in vicinity of the avoided crossing (dashed line), but on the other hand allows for subsequent relaxation even for not-too-high $v$. As a result there is no region in which $Q\sim Q_{\text{SEAL}} \sim 1/v$, however as long as $Q_{\text{SEAL}} \leq 0.1$, i.e. the probability of excitation-relaxation-excitation sequence is relatively low ($<Q_{\text{SEAL}}^2$), the result can be well approximated by taking into account only relaxation processes modifying the excitation generated at the anticrossing, i.e.~$Q \! \approx \! Q_{\text{HEAL}}$. Finally, when the electron transfer time is long enough to allow for a second transition from ground to excited state, i.e.~when $Q_{\text{SEAL}} \geq 0.1$, the HEAL formula gives only a lower bound for results of the AME, as visible at low $v$ when comparing the squares and solid lines.

\begin{figure}[tb]
\includegraphics[width=\columnwidth]{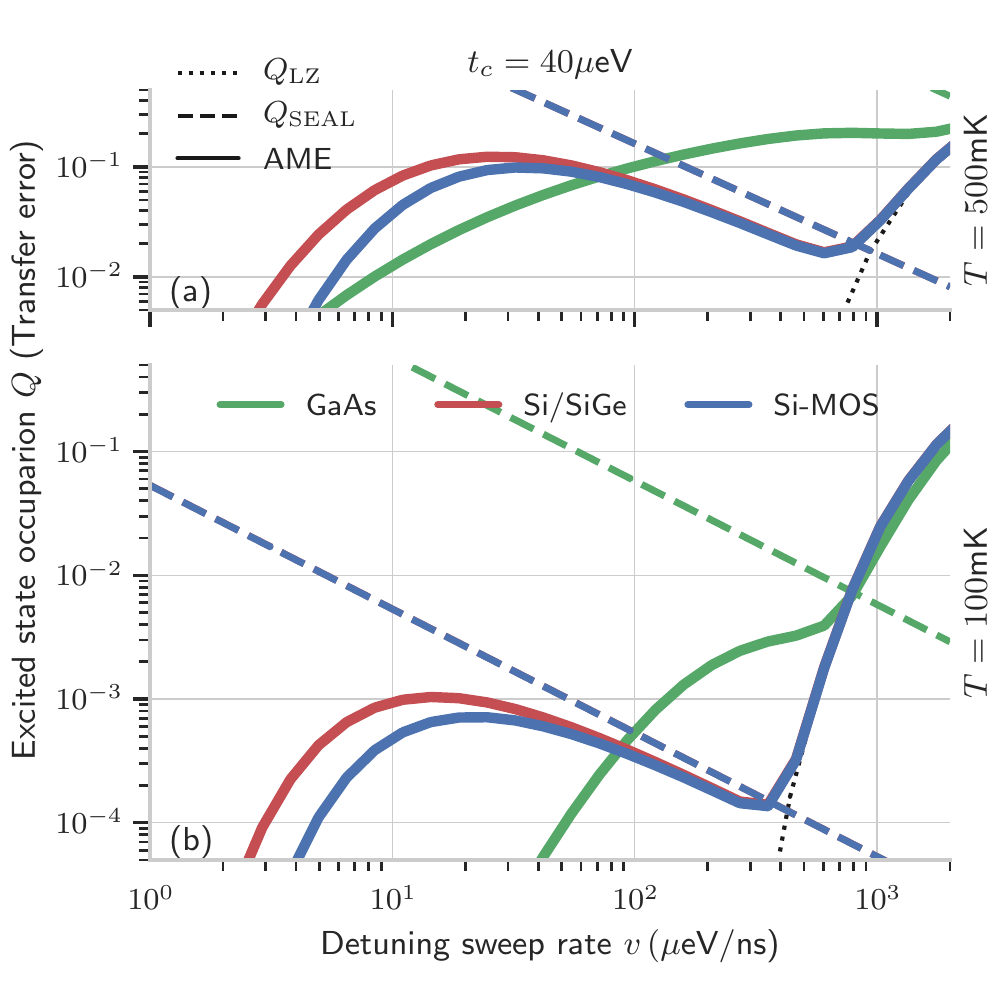}
\caption{Probability of leaving the electron behind in the case of high tunnel coupling $t_c =40\mu$eV. We compare results for $T = 500$mK (a) and $T =100$mK (b), since large tunnel coupling in general should allow for relatively efficient transfer in higher temperature. }
\label{fig:SiGe_GaAs2}
\end{figure} 

\begin{figure}[tbh]
\includegraphics[width=\columnwidth]{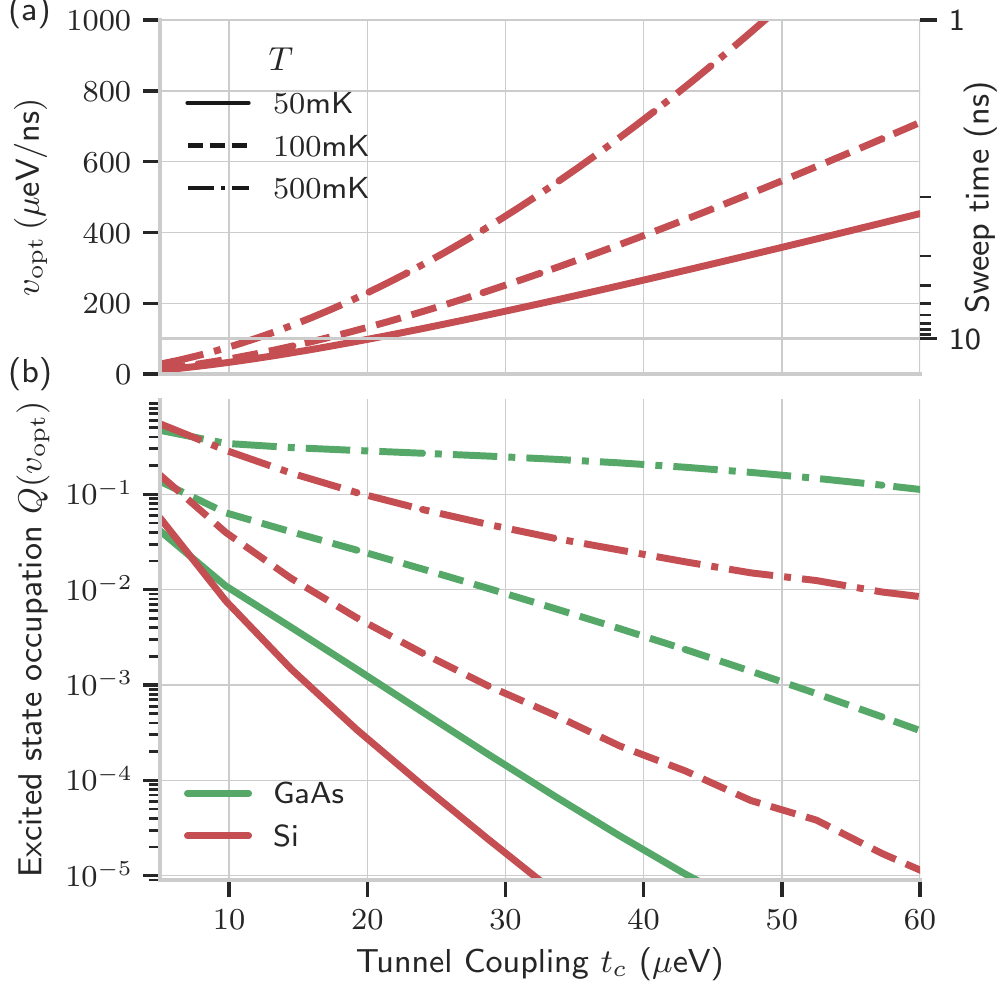}
\caption{Optimal transfer in Si. In panel (a) we plot optimal sweep rate for Si/SiGe and Si-MOS devices $v_{\text{opt}}$ obtained as a solution to Eq.~\eqref{eq:vopt} for $T = 50,100,500$mK (solid, dashed and dotted-dashed lines). In panel (b) we plot probability of leaving the electron behind after the detuning sweep with a rate $v_{\text{opt}}$, as a function of tunnel coupling $t_c$ and for the same selection of temperatures. We compare results for Si against phonon dominated transfer with the same sweep rate in GaAs (Green) as a reference.}
\label{fig:SiGe_GaAs_tc}
\end{figure}

An obvious way to increase the efficiency of charge transfer, or equivalently decrease $Q$, would be to bring the $Q_\text{SEAL}$ result down, as for $\Gamma_+ < \Gamma_-$ it gives an upper bound of excited state occupation induced by environmental fluctuations, i.e. $Q<Q_{\text{SEAL}}$ in the adiabatic regime where $Q>Q_{\text{LZ}}$. This can be achieved by lowering the temperature or increasing the tunnel coupling. In Fig.~\ref{fig:SiGe_GaAs2} we show a rather optimistic result of probability of leaving the electron in the left dot evaluated for the largest $\tun$ reported in the array of Si/SiGe quantum dots \cite{Mills_NC19}, $t_c = 40\mu$eV. As a reference we compare it to the other materials considered, and plot results for $T = 100, 500$mK, as larger tunnel couplings should in principle allow for working at higher temperatures \cite{onoHightemperatureOperationSilicon2019,petitHighfidelityTwoqubitGates2020b,yangOperationSiliconQuantum2020}. We stress that a calculation for $T=50$mK (not shown) gives $Q\leq 10^{-6}$ for $v < 400\mu$eV/ns. For Si nanostructures the behaviour at higher temperatures is qualitatively similar to than shown in Fig.~\ref{fig:SiGe_GaAs}, with a local minimum of $Q = 10^{-4}$,$10^{-2}$ at $v = 400\mu$eV/ns, $800\mu$eV/ns for $T = 100$mK, $500$mK respectively. In GaAs the large value of $\tun$ results in strong coupling between transferred electron and the environment, which at higher temperatures causes flattening of $Q$ as a function of $v$. This effect can be attributed to reaching thermal equilibrium of $Q_{\text{eq}}(\epsilon=0) = \Gamma_+(0)/(\Gamma_+(0)+\Gamma_-(0)) \sim 0.3$ around the avoided crossing, followed by slower relaxation at larger $\epsilon$.

A local minimum of $Q(v)$ is thus expected in both Si-based nanostructures considered here. The value of $Q$ at this minimum can be estimated as the intersection of $Q_{\text{SEAL}}$ and $Q_{\text{LZ}}$, i.e.
\begin{equation}
\label{eq:vopt}
    Q_\text{SEAL}(v_{\text{opt}}) = Q_{\text{LZ}}(v_{\text{opt}}),
\end{equation}
the solution to which is expressed in terms of Lambert $W$ function \cite{abramowitz+stegun} as
$v_{\text{opt}} = \pi t_c^2/2W(a) $
where $W(a)$ satisfies equation $W(a) e^{W(a)} = a$ for $a  = \tfrac{\pi}{4} \sqrt{\beta t_c^{3}}/\Gamma_+(0)$. Since typically $a\gg1$, the asymptotic form of $W(a) = \ln(a) - \ln(\ln(a))$ in the low temperature limit $\beta t_c > 1$ allows to write $v_{\text{opt}} \sim \pi k_{\text{B}}T t_c/2$. As the value of $Q_\text{SEAL}$ is independent of coupling to environment at larger detunings, the sweep rate $v_{\text{opt}}$  which minimizes $Q$ depends on the charge noise amplitude at $\tun$, which is  assumed here to be the same in Si/SiGe and Si-MOS.

In Fig.~\ref{fig:SiGe_GaAs_tc}a we show how $v_\text{opt}$ in Si varies with $t_c$ for $T = 50,100,500$ mK. We see that $v_\text{opt}$ increases as the $Q_{\text{LZ}}$ curve shifts to higher $v$ (due to an increase of $t_c$), or noise-induced excitations $Q_{\text{SEAL}}$ become stronger (here due to an increase of $T$). Next, in Fig.~\ref{fig:SiGe_GaAs_tc}b we use $v_{\text{opt}}$ to compare the corresponding transfer error in Si $Q(v_{\text{opt}})$ (red) against analogous quantity in GaAs (green), as a function of $t_c \in 5 - 60\mu$eV. 
In the Figure we have put together results of the AME (solid, dashed, dotted dashed lines) for three different temperatures $T = 50,100,500$ mK. The probability of losing the electron $Q$ in SiGe appears to be below the value for GaAs for the tunnel couplings apart from the smallest ones of $\tun \sim 5-10\mu$eV, where the corresponding optimal sweep rate ($v_{\text{opt}} \sim 10-50\mu$eV$/$ns depending on the temperature) is large enough to make the Landau-Zener $Q_{LZ}$ probability stay above the phonon-induced $Q$. Of course, $Q$ in GaAs can be made lower by using $v\! < \! v_{\text{opt}}$, but there are other factors that are limiting $v$ from below in GaAs devices (see discussion in the next Section). Similarly, in Si quantum dots charge transfer can be in principle improved by going to much lower sweep rates $v<1\mu$ev/ns, however it would make the few-nanosecond transfer impossible as it has been demonstrated by showing the sweep time interval on the right y-axis of Fig.~\ref{fig:SiGe_GaAs_tc}a.

\begin{figure}[tb]
\includegraphics[width=\columnwidth]{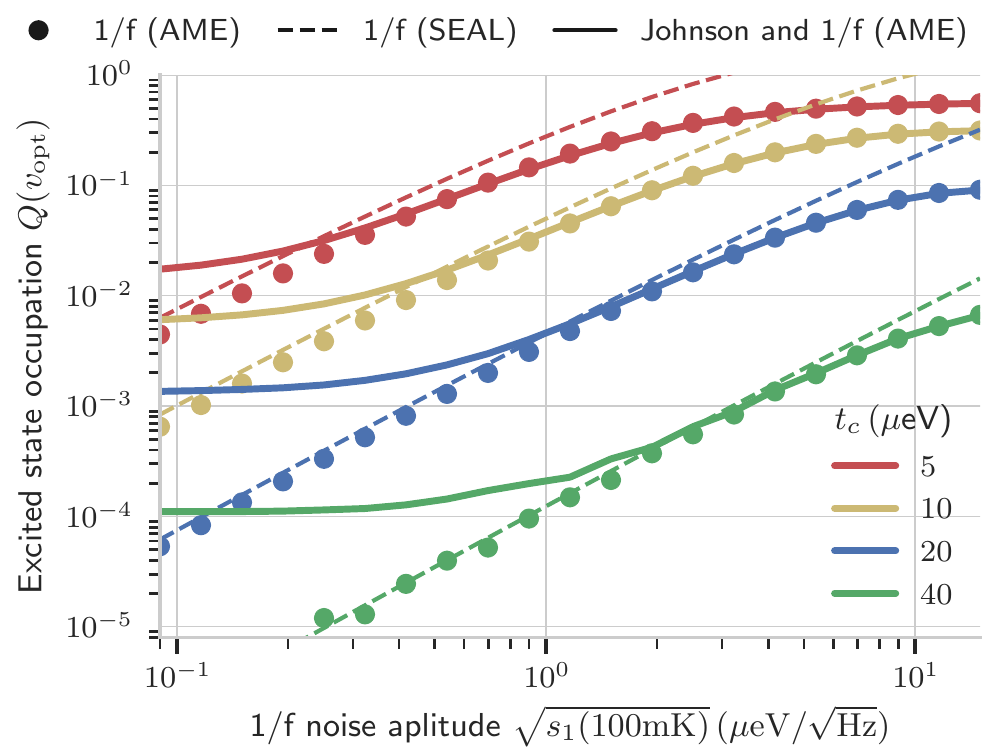}
\caption{Probability of leaving the electron behind using optimal $v_\text{opt}$ sweep rate, as a function of 1/f noise spectral density measured at $1$ Hz and selection of tunnel couplings $t_c = 5,10,20,40\mu$eV at $T = 100$mK. For assumed in Sec.~\ref{sec:charge_noise} model of high-frequency 1/f noise we compare results of adiabatic master equation with (solid line) and without (dots) additional contribution from Johnson noise against $Q_\text{SEAL}$ approximation with 1/f noise only (dashed lines). For the noise in tunnel couplings we assumed $s_1^t = (0.1)^2 s_1^\epsilon$. }
\label{fig:s1}
\end{figure}

The value of optimal sweep rate and corresponding minimum of transfer error $Q(v_\text{opt})$ in Si depends on the amplitude of charge noise at frequency corresponding to tunnel coupling (which is in the GHz range), where its influence dominates over that of phonons. We concentrate here on the amplitude of $1/f$ noise, the amplitude of which can vary by at least an order of magnitude between different Si DQD devices.
In Fig.~\ref{fig:s1} we plot a minimal transfer error at typical electron temperature $T = 100$mK as a function of square root of $1/f$ spectral density evaluated at $1$Hz and at $T=100$mK, which we have previously taken as constant $\sqrt{s_1(100\text{mK})} = 1\mu$eV/$\sqrt{\text{Hz}}$ (see Sec.~\ref{sec:charge_noise}). We plot the results for range of $\tun$ considered here, and emphasize that noise amplitude can be directly related to excitation rate at $\Omega = t_c$ with the following formula:
\begin{equation}
    \Gamma_+^{(1/f)}(t_c) \big[1/\text{ns}\big] \approx  \frac{2s_1 [\mu\text{eV}^2/\text{Hz}]}{t_c [\mu\text{eV}]} \exp(-\frac{t_c}{k_{\text{B}}T})\, \, ,
\end{equation}
where $s_1 = s_1(100\text{mK})$ for brevity and square brackets denoted units in which the quantities should be substituted. The excitation rate obtained using this Equation can be directly used in the SEAL formula, given by Eq.~\eqref{eq:seal}, the result of which was illustrated in Fig.~\ref{fig:s1} using dashed lines. As expected, $Q_{\text{SEAL}}$ agrees well with the results of adiabatic Master equation (dots) for relatively small error $Q(v_\text{opt})\! \ll \!1$. Next we analyze transition between $1/f$ and Johnson noise dominated excitations. The latter can be seen in Fig.~\ref{fig:s1} as a flattening of the solid lines, which represents results of adiabatic master equation with both $1/f$ and Johnson noise, from $R = 50\Omega$ resistor, contributions. By comparing solid and dashed lines, we conclude that amplitude of $1/f$ noise at which it starts to dominate over Johnson noise becomes larger when the tunnel coupling is increased, which can be  deduced from  scaling of respective excitation rates, i.e. $\Gamma_+^{(1/f)}(t_c) \propto 1/t_c$ and $\Gamma_+^{(1/f)}(t_c)^{(\text{J})} \propto t_c$ for $t_c\gg\text{k}_B T$. As the optimal sweep rate $v_{\text{opt}}$ is too high to allow for any phonon-mediated suppression of $Q$ in Si DQDs, the difference between AME and SEAL visible for large noise amplitude is attributed to subsequent relaxation (and further transitions) caused by $1/f$ noise of either large amplitude ($t_c\geq 10\mu$eV) or at relatively high temperature ($t_c =5\mu$eV, for which $\beta t_c \sim 0.5$).

\section{Discussion and summary}
We have presented a theory of the dynamics of a system undergoing a Landau-Zener transition in presence of weak transverse and longitudinal couplings to thermal environments: sources of noise of $1/f$ and Johnson type, and a bath of noninteracting bosons, specifically acoustic vibrations of a three-dimensional crystal. Our focus was on the regime in which the deterministic change of parameters of the Hamiltonian is slow enough to neglect the Landau-Zener  coherent excitation, and the effectively nonadiabatic character of the evolution can be caused only by interaction with the environment. A general theory based on adiabatic Master equation (AME)  was then applied to a case of electron transfer between a pair of voltage-controlled semiconductor quantum dots, for which we took into account realistic parameters for electron-phonon interaction and both Johnson and $1/f$ charge noise. We have calculated transition rates between system's eigenstates as function of interdot detuning $\epsilon$, and used them in AME calculation to obtain the probability of failure of charge transfer between the two dots, $Q$, as a function of detuning sweep rate $v$. 

When $v$ is below the value at which  the Landa-Zener transition is activated, only a finite temperature of environment allows for energy absorption necessary for modification of $Q$, since otherwise electron would stay in the ground state. This absorption most likely takes place in the  vicinity of the anticrossing, where the thermal energy needed for transition is the smallest. A specific feature of the system under consideration is that the dominant coupling to the environment is most effective at the anticrossing, making this effect even stronger. Consequently during the process of electron transfer caused by sweeping the detuning, a finite Q is generated at the anticrossing, when $|\epsilon| \! \leq \! \tun$. Then, for larger positive detunings the electron relaxation processes dominate over the excitation processes, and suppression of $Q$ is expected. 

In the considered DQDs there are two possible scenarios. In Si-based dots, coupling to charge noise dominates, and the transition timescale are longer than the typical transfer times, so that the final $Q$ is very close to the value generated near the anticrossing, which is $\propto \! 1/v$ (proportional to the time spent neart the anticrossing), so it exhibits a dependence on $v$ qualitatively opposite to the one for Landau-Zener effect dominating at large $v$. Only at lowest $v$ the energy relaxation starts to be efficient at lowering $Q$, with this effect being stronger in SiMOS compared to Si/SiGe structures. The competition between the environment-induced excitation and Landau-Zener effect leads then to appearance of optimal $v$, at which $Q$ is minimal.
In GaAs, on the other hand a strong piezoelectric coupling to phonons dominates, transition timescales are shorter than the charge transfer time and consequently many transitions take place, and the final $Q$ monotonically decreases with decreasing $v$, approaching a value exponentially small in final  $\epsilon$, reflecting approaching a thermal occupation of the ground state. 

The main qualitative theoretical result of the paper, which could also apply to systems other than double quantum dots, is thus that a system described by a Landau-Zener model, when coupled  to a thermal environment can realize two possible scenarios: one qualitatively similar to the Landau-Zener effect, but with dependence of $Q$ on $v$ renormalized by environment, and another in which dependence of $Q$ on $v$ is nomonotonic, and there is an optimal sweep rate that minimizes $Q$. 
The main conclusion specific to the considered case of GaAs and Si-based quantum dots is that for $T\! \approx \! 50$ mK,  in GaAs case  $Q$ can be made smaller than $10^{-4}$ by choosing $v$ smaller than $\approx \! 10$  $(100)$ $\mu$eV/ns for $\tun\! =\! 10$ $(20)$ $\mu$eV, while in case of Si having $Q\! =\! 10^{-4}$ requires $\tun \! > \! 20$ $\mu$eV, and optimal $v$ of a few tens of $\mu$eV/ns. Large tunnel couplings and low temperatures are crucial for having small $Q$. In Si-based DQDs there is a possibility of further suppression of $Q$ by decreasing the level of charge noise at GHz frequencies, corresponding to $\tun \! \approx \! 10$ $\mu$eV energy splitting at the anticrossing. 

A process of a controlled electron transfer between two quantum dots is relevant for ongoing attempts at construction of quantum buses based on chains of many tunnel-coupled  dots \cite{Fujita_NPJQI17,Mills_NC19,yonedaCoherentSpinQubitc,Buonascorsi_PRB20,SiQuBus}. 
Let us now discuss the implications of the results of this paper for prospects of coherent shuttling of electron-based spin qubits across $N \! \approx \! 100$ quantum dots. This number of dots in a 1D chain is motivated by requirement of having $\approx \! 10$ $\mu$m distance between few-qubit registers in a realistic architecture of a quantum computer based on gate-controlled QDs \cite{Vandersypen_NPJQI17} and typical interdot separation $\lesssim 100$ nm. 

When the goal of charge shuttling is an on-demand transfer of qubits, which should be highly coherent, and which are to take part in further coherent manipulations after being moved from one register to another, the deterministic character of the shuttling is necessary. Any randomness in qubit arrival times will complicate the application of subsequent coherent operations involving that qubit. Furthermore, any stochastic component in the duration of the qubit transfer will introduce a random contribution to the phase of the qubit. More in-depth discussion of relationship between the indeterministic character of electron shuttling and spin qubit dephasing will be given in \cite{SiQuBus}; here it is enough to realize that large probability of electron arriving at the end of $N$-dot chain at a time other than the desired one, will cause major problems in the context of quantum information processing, and we will treat it here as an error probability. 
Assuming that $Q \! \ll \! 1/N$, the probability of the electron arriving at the end of the chain {\it not} at the desired time, i.e~the probability of qubit transfer-associated error, is $Q_{N}\approx NQ$. 
 
Our results for Si-based quantum dots show that for tunnel couplings in $5-40\mu$eV range, as recently reported in first experimental realizations of electron shuttling over a few dots, achieving $Q_N \! \approx \! 10^{-3}$ will be possible for $\tun\! \geq \! 30$ $\mu$eV and at $T\! \leq \! 50 $ mK. Note that a high-fidelity charge transfer between two dots in Si MOS structure was demonstrated experimentally using $t_c = 450\mu$eV \cite{yonedaCoherentSpinQubitc}, but maintaining such a strong tunnel coupling in a 1D array of $N\! \approx \! 100$ quantum dots will be challenging. 

In GaAs, on the other hand,  $Q_N$ can be made much smaller by decreasing the detuning sweep rate $v$, so that the time of interdot transfer becomes longer than $\sim \! 10$ ns. This in fact also holds for Si-based dots, only $v$ has to be made at least a further order of magnitude smaller. However, for such slow transfers one has to start worrying about well-known mechanisms of spin dephasing that affect the coherence of a static electron localized in a QD. In both the considered materials interaction with nuclei leads to dephasing $T_{2}^{*}$ time of the order or $10$ ns for GaAs \cite{Chekhovich_NM13} and a few hundreds of ns for natural Si \cite{Assali_PRB11,Maune_Nature12,Kawakami_NN14} (and up to tens of microsecond for isotopically purified silicon with about $10^3$ ppm of spinful $^{29}$Si \cite{Assali_PRB11,Yoneda_NN18, StruckNPJQ2020}). For isotopically purified Si QDs in vicinity of micromagnets, their spatially inhomogeneous magnetic fields together with charge noise lead to $T_{2}^{*} \! \approx \! 20$ $\mu$s. Let us use now $T_{2}^*\! =\! 10$ ns ($10$ $\mu$s) for GaAs and Si. In order to avoid significant spin dephasing during the interdot charge transfer, the time of the latter has to be much shorter than $T_2^*$. Assuming that the range of detuning sweep corresponding to the transfer is $\sim \! 1$ meV, the sweep rates have to fulfill $v \! \gg \! 0.1$ $\mu$eV/ns for Si, and $v \! \! \gg 100$ $\mu$eV/ns in GaAs.  In Fig.~\ref{fig:SiGe_GaAs} we see that it means that in GaAs this lower bound on $v$ severely restricts the possibility of lowering $Q$ by making the transfer slower, and in fact a tradeoff between amount of spin dephasing and a finite value of $Q$ due to Landau-Zener effect that dominates the behavior of $Q(v)$ for $v \! \geq \! 100$ $\mu$eV/ns has to be made. In silicon, the lower bound on $v$ is much smaller than $v_{opt}$, so a local minimum of $Q$ visible in the Figure is attainable - but the viability of strategy of lowering $Q$ by using $v \! < \! 1$ $\mu$eV/ns depends on the efficiency of electron relaxation due to charge noise and electron-phonon relaxation (compare red and blue lines, corresponding to Si/SiGe and SiMOS in the Figure) and an exact value of $T_2^*$. All these observations suggest that from the point of view of coherent transfer of a spin qubit,  Si-based quantum dot architectures could have an advantage over GaAs-based ones. 

Let us finish by stressing the main message following from our calculations for realistic GaAs and Si-based quantum dots: the dynamics of inter-dot electron transfer is very strongly affected by electron's interaction with charge noise in Si-based systems and phonons in case of GaAs-based ones. Effects of energy exchange with these environments have to be taken into account to correctly describe the basic physics of electron transfer in the currently available devices. More subtle effects appearing in closed-system description, associated with spin-orbit and valley-orbit (in case of Si) interactions, could become relevant if levels of charge noise are significantly suppressed compared to the currently encountered ones. 

\acknowledgements
We would like to thank Lars Schreiber and Lieven Vandersypen for discussions that motivated us to focus on the problem addressed in the paper, and Piotr Sza\'nkowski for multiple comments on earlier versions of this manuscript. 
This work has been funded by the National Science Centre
(NCN), Poland under QuantERA programme, Grant No.
2017/25/Z/ST3/03044. This project has received funding
from the QuantERA Programme under the acronym Si
QuBus.

\appendix

\section{Correction due to dynamics of classical noise}
\label{app:stationary_phase}
Here we provide detailed calculations of occupation of excited state in the classical limit, i.e. where the fluctuations of detuning and tunnel coupling can be modeled by stochastic contribution to Hamiltonian \eqref{eq:ham0}, i.e. 
\begin{equation}
    \hat H = \frac{\epsilon + \delta \epsilon}{2}\hat \sigma_z + \frac{t + \delta t}{2}\hat \sigma_x.
\end{equation}
As pointed out in the main text, in the limit of weak noise corrections comes from noise dynamics, which in the adiabatic frame modifies off-diagonal element of Hamiltonian \eqref{eq:ham_ad}, written explicitly as
\begin{equation}
    \dot \theta \approx \frac{\sin \theta \delta \dot \epsilon +\cos \theta \delta \dot t  }{\Omega_0},
\end{equation}
where $\Omega_0 = \sqrt{\epsilon^2 + t_c^2}$, $\cos\theta = -\epsilon/\Omega_0$, $\sin\theta = t_c/\Omega_0$ and $\delta \dot \epsilon = \partial_\tau \delta \epsilon$.
\subsection{Leading order perturbation theory}
We evaluate the leading order excitation probability $Q^{(1)}$ due to the noisy term. We use first order time-dependent perturbation theory in adiabatic basis $|\tilde \psi(\tau)\rangle = a_-(\tau) \ket{-,\theta}+ a_+(\tau) \ket{+,\theta}$, and assuming $a_+ = \lambda a_+^{(1)} + \lambda^2 a_+^{(2)} + \ldots $, we compute leading order the correction to occupation of excited state as $Q^{(1)} = \langle|a_+^{(1)}(\tau)|^2\rangle$, which equals
\begin{equation}
\label{eq:cp1}
Q^{(1)} = \frac{1}{4}\int_{-\infty}^{\infty} \Big\langle  \dot \theta(\tau_1) \dot \theta(\tau_2) \Big\rangle\, e^{i\int_{\tau_2}^{\tau_1}\Omega_0(\tau')\text{d}\tau'} \text{d}\tau_1 \text{d}\tau_2,
\end{equation}
where $\langle \ldots \rangle$ denotes classical averaging over noise realizations. The substitution of Eq.~\eqref{eq:dth_der} into Eq.~\eqref{eq:cp1} results in four distinct contributions: 
\begin{equation}
Q^{(1)} = Q^{(1)}_{\epsilon \epsilon} + Q^{(1)}_{tt} + Q^{(1)}_{\epsilon t} + Q^{(1)}_{t\epsilon},
\end{equation}
which correspond to auto- or cross-correlation function of respective noise derivative $Q_{xy}^{(1)} \sim \langle {\delta  \dot  x}(\tau_1) {\delta  \dot y}(\tau_2) \rangle$, where $x,y = \epsilon$ or $t$. For assumed here stationary noises, it is convenient to use the Fourier transform of correlation function, which for the noise derrivative can be expressed in terms of power spectral density of noise (PSD) $S_{xy}(\omega) = \int_{-\infty}^{\infty} \langle\delta x(\tau) \delta y(0) \rangle e^{-i\omega \tau} \text{d}\tau$, i.e. $S_{\dot x\dot y}(\omega) \equiv \omega^2 S_{xy}(\omega)$ \cite{huangSpinRelaxationDue2014}. This allows us to write the correction as
\begin{align}
\label{eq:single_contr}
Q_{xy}^{(1)} =  \int_{-\infty}^{\infty} \frac{\text{d}\omega}{8\pi} S_{xy}(\omega)\, F_x(\omega) F_y^*(\omega),
\end{align}
in which we introduced filtering function $F_x(\omega)$ using Eq.~\eqref{eq:cp1}:
\begin{equation}
\label{eq:fxom}
F_x(\omega) = \int_{-\infty}^{\infty} \text{d}\tau \,f_{x}(\tau)\frac{\omega}{\Omega(\tau)}\,
 \exp{i\omega\tau + i\int_{0}^{\tau} \Omega(\tau')\text{d}\tau'},
\end{equation}
with $f_\epsilon(\tau) = \sin\theta = t_c/\Omega$ and $f_t(\tau)= \cos\theta = -v\tau/\Omega$. 
\subsection{Stationary Phase approximation}
We evaluate the integral \eqref{eq:fxom} for $x = \epsilon, t$, in leading order of stationary phase approximation, where we seek for time at which the argument of exponent $\varphi(\tau) = \omega \tau + \int_0^\tau \Omega(\tau') \text{d}\tau$ is stationary, i.e. $\partial_\tau \varphi(\tau) = 0$, from that $\omega = - \Omega(\tilde \tau)$, which takes place at $ \pm \tilde\tau = \sqrt{(\omega^2 - t_c^2)/v^2}$. Additionally since $\Omega \geq t_c$, the $\omega$ is strictly negative and smaller then $-t_c$. The second derivative, of the phase evaluated at $\tilde \tau$ reads $ \partial_\tau^2 \varphi(\tau)|_{\tau = \tilde \tau} = \pm v \sqrt{1 - t_c^2/\omega^2}$. In the leading order, the integral \eqref{eq:fxom} reads:
\begin{align}
    F_\epsilon \approx &\frac{t_c}{\omega}  \int_{-\infty}^{\infty} \exp{i\varphi(\tilde \tau)+ i\frac{v}{2}\sqrt{1 - t_c^2/\omega^2}(\tau - \tilde \tau)^2} +  \nonumber \\ &+\exp{i\varphi(-\tilde \tau)-i\frac{v}{2}\sqrt{1 - t_c^2/\omega^2}(\tau + \tilde \tau)^2} \text{d}\tau\\
    F_t \approx &\frac{-v\tilde \tau}{\omega}  \int_{-\infty}^{\infty}\exp{i\varphi(\tilde \tau)+ i\frac{v}{2}\sqrt{1 - t_c^2/\omega^2}(\tau - \tilde \tau)^2} +  \nonumber \\ &-\exp{i\varphi(-\tilde \tau)-i\frac{v}{2}\sqrt{1 - t_c^2/\omega^2}(\tau + \tilde \tau)^2}\text{d}\tau
\end{align}
Now we perform Gaussian integration, $\int \text{d}x e^{i a x^2} = \sqrt{\frac{\pi}{ia}}$, using which integrand terms differ by a phase $\sqrt{1/\pm i } = e^{\mp \pi/4}$. Since $\varphi(\tilde \tau) = -\varphi(-\tilde \tau)$ the result can be written as
\begin{align}
    F_\epsilon(\omega) &= \frac{2 t_0 \cos(\varphi(\tilde \tau)-\pi/4)} { \omega} \sqrt{\frac{2\pi}{v}}\left(1 - \frac{t_c^2}{\omega^2}\right)^{-1/4}, \nonumber \\
     F_t(\omega) &= \frac{-2i v\tilde \tau \sin(\varphi(\tilde \tau)-\pi/4)} {\omega} \sqrt{\frac{2\pi}{v}}\left(1 - \frac{t_c^2}{\omega^2}\right)^{-1/4}
\end{align}

First we consider diagonal part ($x=y$) of  Eq.~\eqref{eq:single_contr}, in which $|F_\epsilon|^2 \propto \cos^2(\varphi - \pi/4)$ and $|F_t|^2\propto \sin^2(\varphi - \pi/4)$. Due to rapidly oscillating nature of both functions, we replace them by average values of $\cos^2\varphi$ and $\sin^2\varphi$ equal 1/2, which leads to:
\begin{align}
Q_{\epsilon\epsilon}^{(1)} &= \frac{1}{2v} \int_{-\infty}^{-t_c} \text{d}\omega S(\omega) \frac{t_c^2}{\omega^2} \left(1 - \frac{t_c^2}{\omega^2}\right)^{-1/2} \nonumber \\
Q_{tt}^{(1)} &= \frac{1}{2v} \int_{-\infty}^{-t_c} \text{d}\omega S(\omega) \frac{\omega^2 - t_c^2}{\omega^2} \left(1 - \frac{t_c^2}{\omega^2}\right)^{-1/2},
\end{align}
where the strictly negative value of $\omega$, reflects absorption of energy quanta. Finally we conclude by showing that cross-correlation is negligibly small. We use the argument that $F_t^* = -F_t$ is strictly imaginary, and as a result we have
\begin{equation}
    Q_{\epsilon t}^{(1)} +  Q_{t\epsilon }^{(1)} = \int_{-\infty}^{\infty} (S_{\epsilon t}(\omega) - S_{t\epsilon}(\omega)) F_\epsilon(\omega) F_t^*(\omega),
\end{equation}
where the integrand is equivalent to imaginary part of cross-spectrum, and hence vanishes for $\langle \delta \epsilon(\tau) \delta t\rangle = \langle \delta t \delta \epsilon\rangle$. Non-trivial imaginary part of cross-spectrum results only from causal relation between $\delta t$, $\delta \epsilon$ \cite{Szankowski2016}, however even in such special case we argue that $F_\epsilon(\omega) F_t^*(\omega) \propto \cos(\varphi + \pi/4) \sin(\varphi + \pi/4)$ which due to zero average is expected to be much smaller than auto-correlation contributions. As a result corrections to occupation of excited state due to weak classical noise can be written as:
\begin{align}
\label{eq:corrections}
Q_{\epsilon\epsilon}^{(1)} &= \frac{1}{2v} \int_{t_c}^{\infty}\text{d}\Omega \frac{ S_{\epsilon}(-\Omega)}{\sqrt{1 - t^2/\Omega^2}}\left(\frac{t_c^2}{\Omega^2}\right)   \nonumber \\
Q_{ tt}^{(1)} &=  \frac{1}{2v} \int_{t_c}^{\infty}\text{d}\Omega\,S_{ t}(-\Omega)\sqrt{1 - \frac{t_c^2}{\Omega^2}},
\end{align}
using which we recovered high frequency limit of \cite{Malla_PRB17} where lower bound of the integrals reflects the minimal energy needed for the excitation to occur. Due to the dominant role of longitudinal component $\delta \epsilon$ we omit here contributions from frequencies below $t_c$, which are relevant only for transverse $\delta t$ noise \cite{Malla_PRB17, Luo_PRB17}. In particular corrections from quasi-static noise in tunnel coupling vanishes in assumed here weak noise ($\delta t \! \ll \! \tun$) and adiabatic ($t_c^2>v$) limits \cite{Kayanuma_JPSJ85}. 
\begin{figure}[tb!]
\includegraphics[width=\columnwidth]{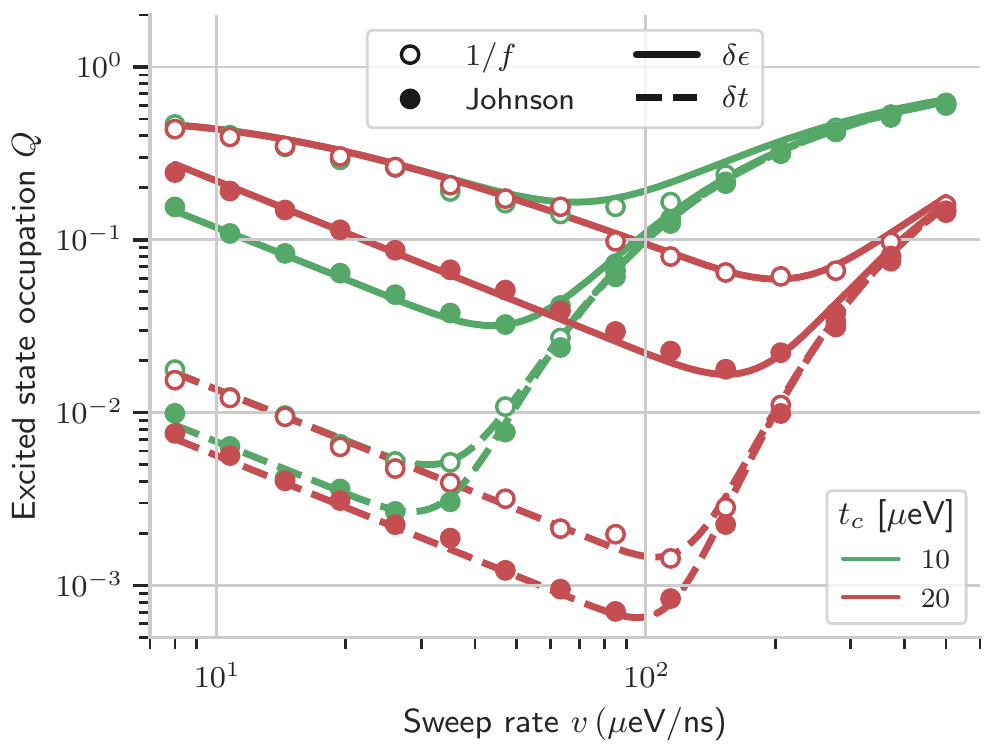}
\caption{Probability of occupying higher energy state after detuning sweep $Q$ in presence of white (filled) and 1/f (hollow dots) classical noise in detuning/tunneling as a function of sweep rate for tunnel couplings $t_c = 10\mu$eV (green) and $20\mu$eV (red). Points correspond to numerical simulation of Schrodinger equation averaged over realization of classical noise process. Lines correspond to analytical expression $Q^{(\infty)}$ with the rates calculated according to Eq.~\eqref{eq:noise_gam} for the noise in detunning $\delta \epsilon$ (solid) and tunnel coupling $\delta t$ (dashed). For illustration we used arbitrary parameters for detuning noise $S^{\epsilon}_{0.1} T/T_{0.1} = (1.5)^2\mu$eV$^2$/Hz (1/f) and $2Jk_\text{B}T = (0.3)^2\mu$eV$^2$/Hz (White part of Johnsons noise). The tunnel coupling fluctuations are reduced by a factor of 10, i.e. $S^{\epsilon}(\omega) = (10^2)S^t(\omega)$. To emulate high temperature limit we set terminal sweep rate to $\epsilon_f = 100\mu$eV, which corresponds of thermal energy at $T \approx 1.2 $K .
}
\label{fig:noise}
\end{figure}
\subsection{Transition rates}
The first order calculation can be interpreted as a probability of single transition from ground to excited state during adiabatic transfer, and as such can be written as an integral of transition rate $Q^{(1)} = \int \text{d} \tau' \Gamma_\infty(\tau')$, see Eq.~\eqref{eq:q1}. An explicit form of $\Gamma_\infty$ can be deduced from Eq.~\eqref{eq:corrections} as
\begin{align}
\label{eq:noise_gam}
\Gamma_{\infty,\epsilon}(\tau)&= \frac{1}{4} \left(\frac{t_c}{\Omega(\tau)}\right)^2\,S_\epsilon^\text{cl}\big(\Omega(\tau)\big)\nonumber \\  
\Gamma_{\infty,t}(\tau) &= \frac{1}{4}\bigg(1-\left(\frac{t_c }{\Omega(\tau)}\right)^2\bigg)\,S_t^\text{cl}\big(\Omega(\tau)\big).
\end{align}
Finally, we prove that a result obtained by substituting $\Gamma_\pm =\Gamma_\infty = \Gamma_{\infty,\epsilon} + \Gamma_{\infty,t}$ into rate equation Eq.~\eqref{eq:rate}, which results in high temperature solution:
\begin{equation}
    Q^{(\infty)} = \frac{1}{2}\left(1 - \exp{-2\int_{\tau_i}^{\tau_f}\Gamma_\infty(\tau') \text{d}\tau'}\right),
\end{equation}
is equivalent to an evolution driven by Hamiltonian Eq.~\eqref{eq:ham0}, averaged over realizations of classical fluctuations of parameters. In Fig.~\ref{fig:noise} we have separately plotted contributions from detuning noise $\delta \epsilon$ (solid line), and tunnel coupling noise $\delta t$ (dashed line), as a result of $1/f$ noise (hollow dots) and white noise (filled dots). Independently of the considered noise type, in the fast sweep rate limit ($v\gg t_c^2$) we recover the Landau-Zener solution, for which $Q_\text{LZ}$ depends on the relation between $v$ and $t_c$ only, and thus for sufficiently large $v$ the results group according to tunnel couplings $t_c=10\mu$eV (green) and $t_c=20\mu$eV (red). In the low sweep rate limit, for $1/f$ noise in detuning (solid line, hollow dots) we obtain results from \cite{Krzywda_PRB20}, for which $Q_\epsilon^{(\infty)}$ ($Q^{(\infty)}$ with $\Gamma_\infty = \Gamma_{\infty,\epsilon}$) is independent of $t_c$. The same applies to $1/f$ noise in tunnel coupling (dashed lines, hollow dots), for which a 10-fold decreased noise amplitude (compared to the case of detuning noise) translates into almost 2 orders of magnitude lower $Q_t^{(\infty)}$. For the white part of Johnson's noise, distinction between different $\tun$ is much more visible for noise in detuning (solid lines, filled dots), since larger $\tun$ significantly increases the time spent in vicinity of avoided crossing $\approx \! 2t_c/v$, during which the longitudinal transitions $\Gamma_{\infty,\epsilon}\propto (t_c/\Omega)^2$ are most effective. In the case of noise in tunnel coupling the opposite is true, since larger $\tun$ only slightly decreases time spent outside of the avoided crossing region, while $\Gamma_{\infty,t} \propto (v\tau/\Omega)^2$.

\section{Details of phonon relaxation rate}
\label{app:phonons}
We now turn to evaluation of zero-temperature phonon relaxation rate in more details. First we show how orbital and interdot phonon-related processes emerge when using the $\ket{\pm,\theta}$ basis of eigenstates of instantaneous Hamiltonian. Next we investigate the elements for Gaussian choice of electron wavefunctions, and discuss the role of harmonic and Hund-Muliken approximation. 
\subsection{Interdot and orbital processes}
We start with the Phonon spectral density, given by Eq.~\eqref{eq:spectrum_ph}, which predicts $S(\omega) \propto |\bra{-} e^{i\mathbf{k}\mathbf{r}} \ket{+}|^2$. We now evaluate the matrix element, by pluging in adiabatic basis given by Eq.~\eqref{eq:states}, which results in:
\begin{align}
\label{eq:app:orbital}
    \bra{-,\theta} e^{i\mathbf{k}\mathbf{r}} &\ket{+,\theta} =  \nonumber \\ &\cos\theta \Re{\bra{L}e^{i\mathbf{k}\mathbf{r}}\ket{R}} +\Im{\bra{L}e^{i\mathbf{k}\mathbf{r}}\ket{R}} \nonumber \\ &+\frac{1}{2}\sin\theta(\bra{L}e^{i\mathbf{k}\mathbf{r}}\ket{L} - \bra{R}e^{i\mathbf{k}\mathbf{r}}\ket{R}),
\end{align}
where $\tan\theta = -\epsilon/t_c$, and consecutive terms corresponds to interdot ($\hat \sigma_x$, $\hat \sigma_y$) and orbital $\hat \sigma_z$ coupling in dots basis respectively. In absence of large magnetic field in $z$ direction, wavefunctions can be assumed real, hence $\Im{\bra{L}e^{i\mathbf{k}\mathbf{r}}\ket{R}} = 0 $. The exact form of matrix element depends on the assumed form of wavefunctions, i.e. $\psi_{L/R}(\mathbf{r}) = \bra{\mathbf{r}}\ket{L/R}$, which will be investigated below.

\subsection{Hund-Mulliken approximation}
 Before invoking concrete form of  wavefunction of an electron localized in a QD, let us  comment on the so called Hund-Mulliken approximation, in  which one uses orthogonalized orbitals $\ket{L/R} = \mathcal N(\ket{L_0/R_0} - g \ket{R_0/L_0})$ built from bare wavefunction of electrons in isolated quantum dots: $\ket{L_0/R_0}$, with  $\mathcal N$ being normalization constant. The parameter $g$ is a function of the overlap $l = \bra{L_0}\ket{R_0} \ll 1$, with its value given by the orthogonality condition:
 \begin{equation}
    \bra{L}\ket{R} = \mathcal N^2 (l - 2g + lg^2)  = 0
 \end{equation}
 from which $g =( 1 - \sqrt{1-l^2})/l \sim l/2$ for $l \ll 1$. Consistently we concentrate on leading order in $g$ or $l$, according to which and $1 =\bra{L}\ket{L} = \mathcal N^2 (1 - 2gl + g^2 )$ we take $\mathcal N \sim 1$. Assuming real wavefunction  $\ket{L_0/R_0}$, we substitute the orthogonalized  states into Eq.~\eqref{eq:app:orbital} from which we obtain:
 \begin{align}
 \label{eq:app:dipole}
     &\bra{-} e^{i\mathbf{k}\mathbf{r}} \ket{+}\sim \nonumber \\&\sim \cos\theta\bigg(\bra{L_0}e^{i\mathbf{k}\mathbf{r}}\ket{R_0} - g \big(\bra{L_0}e^{i\mathbf{k}\mathbf{r}}\ket{L_0} + \bra{R_0}e^{i\mathbf{k}\mathbf{r}}\ket{R_0}\big)\bigg)\nonumber \\
     &+\frac{1}{2}\sin\theta\bigg( \bra{L_0}e^{i\mathbf{k}\mathbf{r}}\ket{L_0}- \bra{R_0}e^{i\mathbf{k}\mathbf{r}}\ket{R_0}\bigg),
 \end{align}
 where in the latter term correction linear in the overlap $g$ cancels.
 
 \subsection{Harmonic approximation}
 Finally we substitute concrete form of isolated wavefunctions, and evaluate matrix element $\bra{-}e^{i\mathbf{k}\mathbf{r}}\ket{+}$. We assume the wavefunction is indepedent in all three directions ($\Psi_{L_0/R_0}(\mathbf{r}) = \psi_{L_0/R_0}(x,y) \psi_z(z)$) and in has a Gaussian shape: 
 \begin{align}
     &\psi_{L_0/R_0}(x,y) = \frac{1}{(\pi^2 r_{xy}^4)^{1/4}} \exp{-\frac{(x\pm \Delta x/2)^2 + y^2}{2r_{xy}^2}} \nonumber \\
     &\psi_{z}(z) = \frac{1}{(\pi r_{z}^2)^{1/4}} \exp{-\frac{z^2}{2r_{z}^2}},
 \end{align}
 such that for electron wavefunction, FWHM$_x \approx 2 r_{xy}$ and FWHM$_z \approx 2r_{z}$. In such case Eq.~\eqref{eq:app:dipole} reads:
 \begin{align}
     \bra{-} e^{i\mathbf{k}\mathbf{r}} \ket{+}&\sim \exp{-\frac{k_{xy}^2r_{xy}^2 + k_z^2 r_{z}^2}{4}}\times \nonumber \\
     \bigg(&\cos\theta \exp{-\frac{\Delta x^2}{4r_{xy}^2}}\big(1 - \cos(k_x \Delta x/2)\big) + \nonumber \\
     &\qquad\qquad -i\sin\theta \sin(k_x \Delta x/2)\bigg),
 \end{align}
 where we used that in harmonic approximation $g = l/2 = \tfrac{1}{2}e^{-\Delta x^2/4r_{xy}^2}$. Interdot and orbital relaxation are given by real and imaginary part of above matrix element and hence cause relaxation independently.


%

\end{document}